\documentclass[aps,pre,floatfix,reprint,groupedaddress,showkeys,superscriptaddress]{revtex4-1}
\usepackage[utf8x]{inputenc}
\usepackage{graphicx}
\usepackage{dcolumn}
\usepackage{bm}
\usepackage{epstopdf}
\usepackage{amsmath}
\usepackage{float}
   \usepackage[all]{xy}
\usepackage{subfigure}
\usepackage{xcolor}
\usepackage{soul} 
\usepackage{wrapfig}

  \graphicspath{{./figure/}}
\newcommand{\vect}[1]{\boldsymbol{#1}}
\renewcommand{\figurename}{Figure}
\renewcommand{\tablename}{Table}
\renewcommand{\thetable}{\arabic{table}}

\newcommand{\beginsupplement}{
        \setcounter{table}{0}
        \renewcommand{\tablename}{Supplementary Table}
        \renewcommand{\thetable}{\arabic{table}}
        \setcounter{figure}{0}
        \renewcommand{\figurename}{Supplementary Figure}
        \renewcommand\appendixname{Supplementary Note}
        \renewcommand{\thesection}{\arabic{section}}
     }

\begin{document}
	
\title{Efficient generative modeling of protein sequences using simple
  autoregressive models}

\author{Jeanne Trinquier}
\affiliation{
Sorbonne Universit\'e, CNRS, Institut de Biologie Paris Seine, Biologie
Computationnelle et Quantitative LCQB, F-75005 Paris, France
}
\affiliation{
Laboratoire de Physique de l'Ecole Normale Sup\'erieure, ENS, Universit\'e PSL, CNRS, Sorbonne Universit\'e, Universit\'e de Paris, F-75005 Paris, France
}

\author{Guido Uguzzoni}
\affiliation{
Department of Applied Science and Technology (DISAT), Politecnico di Torino, Corso Duca degli Abruzzi 24, I-10129 Torino, Italy}
\affiliation{
Italian Institute for Genomic Medicine, IRCCS Candiolo, SP-142, I-10060 Candiolo (TO) - Italy}

\author{Andrea Pagnani}
\affiliation{
Department of Applied Science and Technology (DISAT), Politecnico di Torino, Corso Duca degli Abruzzi 24, I-10129 Torino, Italy}
\affiliation{
Italian Institute for Genomic Medicine, IRCCS Candiolo, SP-142, I-10060 Candiolo (TO) - Italy}
\affiliation{
INFN Sezione di Torino, Via P. Giuria 1, I-10125 Torino, Italy
}

\author{Francesco Zamponi}
\affiliation{
Laboratoire de Physique de l'Ecole Normale Sup\'erieure, ENS,
Universit\'e PSL, CNRS, Sorbonne Universit\'e, Universit\'e de Paris,
F-75005 Paris, France 
}

\author{Martin Weigt}
\affiliation{
Sorbonne Universit\'e, CNRS, Institut de Biologie Paris Seine, Biologie
Computationnelle et Quantitative LCQB, F-75005 Paris, France
}
\affiliation{
Email: martin.weigt@sorbonne-universite.fr
}

\begin{abstract}
\section*{Abstract}
Generative models emerge as promising candidates for novel sequence-data driven approaches to protein design, and for the extraction of structural and functional information about proteins deeply hidden in rapidly growing sequence databases. Here we propose simple autoregressive models as highly accurate but computationally efficient generative sequence models. We show that they perform similarly to existing approaches based on Boltzmann machines or deep generative models, but at a substantially lower computational cost (by a factor between $10^2$ and $10^3$). Furthermore, the simple structure of our models has distinctive mathematical advantages, which translate into an improved applicability in sequence generation and evaluation. Within these models, we can easily estimate both the probability of a given sequence, and, using the model's entropy, the size of the functional sequence space related to a specific protein family. In the example of response regulators, we find a huge number of ca.~$10^{68}$ possible sequences, which nevertheless constitute only the astronomically small fraction $10^{-80}$ of all amino-acid sequences of the same length. These findings illustrate the potential and the difficulty in exploring sequence space via generative sequence models.
\end{abstract}

\maketitle

\section*{Introduction}

The impressive growth of sequence databases is prompted by
increasingly powerful techniques in data-driven modeling,
helping to extract the rich information hidden in raw data. In the
context of protein sequences, unsupervised learning techniques are of
particular interest: only about 0.25\% of the more than 200 million
amino-acid sequences currently available in the Uniprot database \cite{uniprot2019uniprot} have 
manual annotations, which can be used for supervised methods.

Unsupervised methods may benefit from evolutionary relationships
between proteins: while mutations modify amino-acid sequences, 
selection keeps their biological functions and their
three-dimensional structures remarkably conserved. The Pfam protein
family database \cite{el2019pfam}, {\em e.g.}, lists more than 19,000 families of
homologous proteins, offering rich datasets of sequence-diversified
but functionally conserved proteins.

In this context, generative statistical models are rapidly
gaining interest. The natural sequence variability across a protein family
is captured via a probability $P(a_1,...,a_L)$ defined for all amino-acid
sequences $(a_1,...,a_L)$. Sampling from $P(a_1,...,a_L)$ can be used to 
generate new, non-natural amino-acid sequences, which in an ideal case
should be statistically indistinguishable from the natural sequences.
However, the task of learning $P(a_1,...,a_L)$ is highly non-trivial:
the model has to assign probabilities to all $20^L$ 
possible amino-acid sequences. For typical proteins of lengths
$L=50-500$, this accounts to $10^{65}-10^{650}$ values, to be learned from 
the $M=10^3-10^6$ sequences contained in most protein families. Selecting 
adequate generative model architectures is thus of outstanding importance.

The currently best explored generative models for proteins are
so-called coevolutionary models~\cite{de2013emerging}, such as those constructed by the Direct
Coupling Analysis (DCA)~\cite{morcos2011direct,cocco2018inverse,figliuzzi2018pairwise} (a more detailed review of the state of the art is provided below). They explicitly model
the usage of amino acids in single positions ({\em i.e.} residue
conservation) and correlations between pairs of positions ({\em i.e.}
residue coevolution). The resulting models are mathematically equivalent to Potts models~\cite{levy2017potts} in statistical physics, or to Boltzmann machines in statistical learning~\cite{ackley1985learning}. They have found numerous applications in protein biology.

The effect of amino-acid mutations is predicted via the
  log-ratio $\log\{ P({\rm mutant})/P({\rm 
    wildtype})\}$ between mutant and wildtype probabilities.
  Strong correlations to mutational effects determined
  experimentally via deep mutational scanning have been reported~\cite{figliuzzi2016coevolutionary,hopf2017mutation}. Promising application are the data-driven design of mutant libraries for protein optimization~\cite{cheng2014toward,cheng2016connecting,reimer2019structures}, and the use of Potts models as sequence landscapes in quantitative models of protein evolution~\cite{de2020epistatic,bisardi2021modeling}.

Contacts between residues in the protein fold are
  extracted from the strongest epistatic couplings between double
  mutations, {\em i.e.} from the direct couplings giving
  the name to DCA~\cite{morcos2011direct}. These couplings are essential input features in
  the wave of deep-learning (DL) methods, which currently revolutionize the field of protein-structure 
  prediction~\cite{wang2017accurate,greener2019deep,senior2020improved,yang2020improved}.  
  
 The generative implementation bmDCA~\cite{figliuzzi2018pairwise} is able to  generate  artificial but functional amino-acid sequences~\cite{tian2018co,russ2020evolution}. Such observations suggest novel but almost unexplored approaches towards data-driven protein design, which complement current approaches based mostly on large-scale experimental screening of randomized sequence libraries or time-intensive bio-molecular simulation, typically followed by sequence optimization using directed evolution, cf.~\cite{jackel2008protein,huang2016coming} for reviews.

\begin{figure*}[!htb]
	\begin{center}
		\includegraphics[keepaspectratio,width=.75\textwidth]{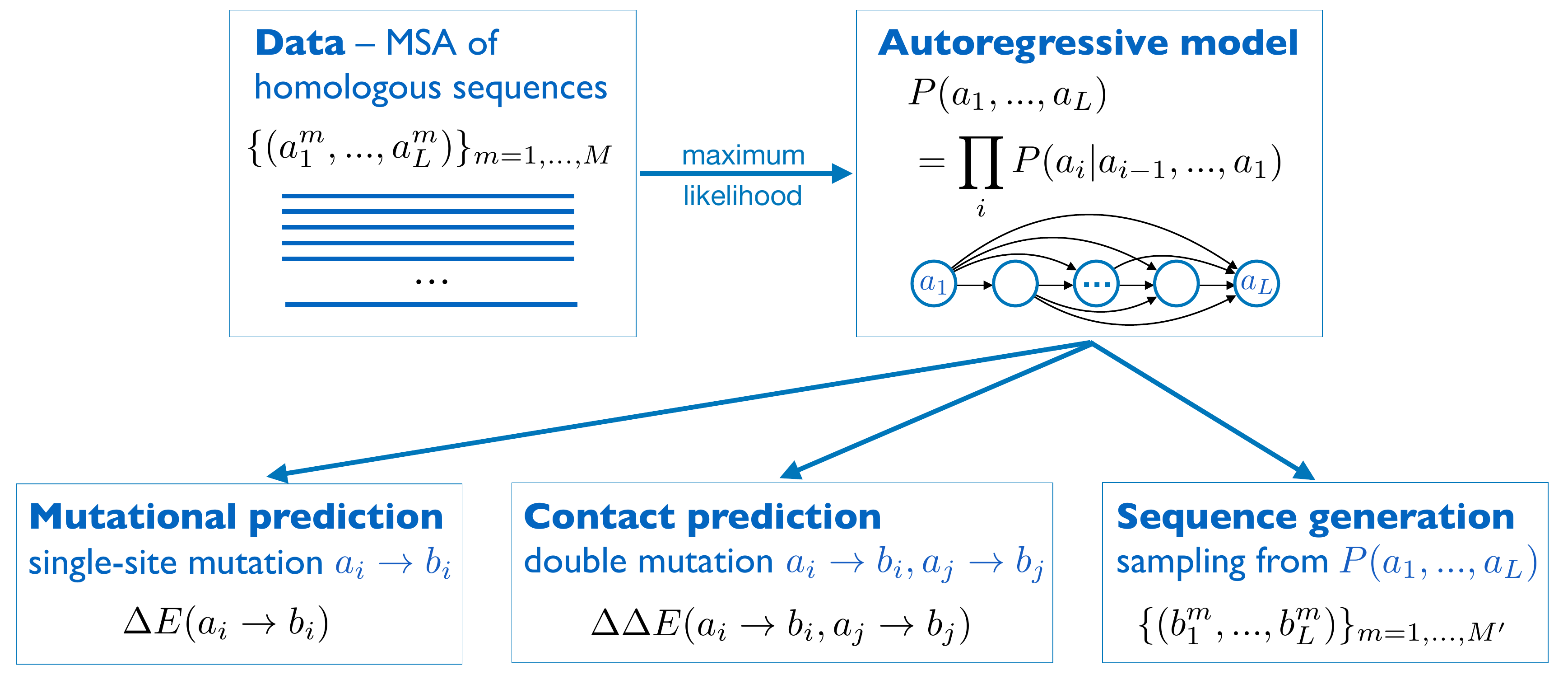}
	\end{center}
	\vspace{-5mm}
	\caption{Schematic representation of the arDCA approach: 
	Starting from a MSA of homologous sequences, we 
	use maximum-likelihood inference to learn an autoregressive 
	model, which factorizes the joint sequence probability
	$P(a_1,...,a_L)$ into conditional single-residue probabilities 
	$P(a_i|a_{i-1},...,a_1)$. Defining the statistical energy
	$E(a_1,...,a_L) = - \log P(a_1,...,a_L)$ of a sequence, we
	consequently predict mutational effects and contacts as statistical 
	energy changes when substituting residues individually or in pairs, 
	and we design new sequences by sampling from $P(a_1,...,a_L)$.}
        \label{fig:graphical_abstract}
\end{figure*}

Here we propose a simple model architecture called arDCA, 
based on a shallow (one-layer) autoregressive model paired with generalized logistic regression. 
Such models are computationally very efficient, they can be
learned in few minutes, as compared to days for bmDCA and more involved
architectures. Nevertheless, we demonstrate that arDCA provides highly
accurate generative models, comparable to the state of the art in
mutational-effect and residue-contact prediction. Their simple
structure makes them more robust in the case of limited data. Furthermore, 
and this may have important applications in homology detection \cite{wilburn2020remote},
our autoregressive models are the only generative models we know about, which allow for calculating exact sequence probabilities, and not only non-normalized sequence weights. Thereby arDCA enables the comparison of the same sequence in different models for different protein families. Last but not least, the entropy of arDCA models, which is related to the size of the functional sequence space associated to a given protein family, can be computed much more efficiently than in bmDCA.

Before proceeding, we provide here a short review of the state of the art in generative protein modeling. The literature is extensive and rapidly growing, so we will concentrate on the methods being most directly relevant as compared to the scope our work. 

 We focus on generative models purely based on sequence data. The sequences belong to homologous protein families, and are given in form of multiple sequence alignments (MSA), {\em i.e.} as a rectangular matrix ${\cal D}=(a_i^m|i=1,...,L; m=1,...,M)$ containing $M$ aligned proteins of length $L$. The entries $a_i^m$ equal either one of the standard 20 amino acids, or the alignment gap ``--''. In total, we have $q=21$ possible different symbols in ${\cal D}$. The aim of unsupervised generative modeling is to learn a statistical model $P(a_1,...,a_L)$ of (aligned) full-length sequences, which faithfully reflects the variability found in ${\cal D}$: sequences belonging to the protein family of interest should have comparably high probabilities, unrelated sequences very small probabilities. Furthermore, a new artificial MSA ${\cal D}'$ sampled sequence by sequence from model $P(a_1,...,a_L)$ should be statistically and functionally indistinguishable from the natural aligned MSA ${\cal D}$ given as input. 

A way to achieve this goal is the above-mentioned use of Boltzmann-machine learning based on conservation and coevolution, which leads to pairwise-interacting Potts models, {\em i.e.}~bmDCA~\cite{figliuzzi2018pairwise},
and related methods~\cite{sutto2015residue,barton2016ace,vorberg2018synthetic}. An alternative implementation of bmDCA, including a decimation of statistically irrelevant couplings, has been presented in~\cite{barrat2020sparse} and is the one used as a benchmark in this work; the Mi3 package~\cite{haldane2021mi3} also provides a GPU-based accelerated implementation.

However, Potts models or Boltzmann machines are not the only generative-model architectures explored for protein sequences. Latent-variable models like Restricted Boltzmann machines~\cite{tubiana2019learning} or Hopfield-Potts models~\cite{shimagaki2019selection} learn dimensionally reduced representations of proteins; using sequence motifs, they are able to capture groups of collectively evolving residues~\cite{rivoire2016evolution} better than DCA models, but are less accurate in extracting structural information from the learning MSA~\cite{shimagaki2019selection}.

An important class of generative models based on latent variables are variational autoencoders (VAE), which achieve dimensional reduction, but in the flexible and powerful setting of deep learning. The DeepSequence implementation~\cite{riesselman2018deep} was originally designed and tested for predicting the effects of mutations around a given wild type. It currently provides one of the best mutational-effect predictors, and we will show below that arDCA provides comparable quality of prediction for this specific task. The DeepSequence code has been modified in~\cite{mcgee2020generative} to explore its capacities in generating artificial sequences being statistically indistinguishable from the natural MSA; it was shown that its performance was substantially less accurate than bmDCA.
Another implementation of a VAE was reported in~\cite{hawkins2020generating}; also in this case the generative performances are inferior to bmDCA, but the organization of latent variables was shown to carry significant information on functionality. Furthermore, some generated mutant sequences were successfully tested experimentally. Interestingly, it was also shown that learning VAE on unaligned sequences decreases the performance as compared to pre-aligned MSA as used by all before-mentioned models. This observation was complemented by Ref.~\cite{costello2019}, which reported a VAE implementation trained on non-aligned sequences from UniProt, with length $10<L<1000$. The VAE had good reconstruction accuracy for small $L<200$, which however dropped significantly for larger $L$. The latent space also in this case shows an interesting organization in terms of function, which was used to generate {\it in silico} proteins with desired properties, but no experimental test was provided. The paper does not report any statistical test of the generative properties (such as a Pearson correlation of two-point correlations), and the publicly not yet available code makes a quantitative comparison to our results currently impossible. 

Another interesting DL architecture is that of a Generative Adversarial Network (GAN), which was explored in~\cite{repecka2021expanding} on a single family of aligned homologous sequences. While the model has a very large number of trainable parameters ($\sim$60M), it seems to reproduce well the statistics of the training MSA, and most importantly, the authors could generate an enzyme with only 66\% identity to the closest natural one, which was still found to be functional {\it in vitro}. An alternative implementation of the same architecture was presented in~\cite{amimeur2020designing}, and applied to the design of antibodies; also in this case the resulting sequences were validated experimentally.

Not all generative models for proteins are based on sequence ensembles. Several research groups explored the possibility of generating sequences with given three-dimensional structure~\cite{ingraham2021generative,anand2020protein,jing2020learning}, e.g. via a VAE~\cite{greener2018} or a  Graph Neural Network~\cite{strokach2020fast}, or by inverting structural prediction models~\cite{norn2021protein,anishchenko2020novo,linder2020fast,fannjiang2020autofocused}. It is important to stress that this is a very different task from ours (our work does not use structure), so it is difficult to perform a direct comparison between our work and these ones. It would be interesting to explore, in a future work, the possibility to unify the different approaches and to use sequence and structure jointly for constructing improved generative models.

In summary, for the specific task of interest here, namely generate an artificial MSA statistically indistinguishable from the natural one, one can take as reference models bmDCA~\cite{figliuzzi2018pairwise,barrat2020sparse} in the context of Potts-model-like architectures, and DeepSequence~\cite{riesselman2018deep} in the context of deep networks. We will show in the following that arDCA performs comparably to bmDCA, and better than DeepSequence, at strongly reduced computational cost.
From anecdotal evidence in the works mentioned above, and in agreement with general
observations in machine learning, it appears that deep architectures may
be more powerful than shallow architectures, provided that very large
datasets and computational resources are available~\cite{riesselman2018deep}. Indeed, we will show that for the related task of single-mutation predictions around a wild type, DeepSequence outperforms arDCA on rich datasets, while the inverse is true on small datasets.

\section*{Results}

\subsection*{Autoregressive models for protein families}

Here we propose a computationally efficient approach
based on autoregressive models. We start from the exact
decomposition 
\begin{equation}
  \label{eq:ar}
  P(a_1,...,a_L) = P(a_1)\cdot P(a_2|a_1)\cdots P(a_L|a_{L-1},...,a_1)\ ,
\end{equation}
of the joint probability of a full-length sequence into
a product of more and more involved conditional probabilities
$P(a_i|a_{i-1},...,a_1)$ of the amino acids $a_i$ in single positions, conditioned to all previously seen positions $a_{i-1},...,a_1$. While this decomposition is a direct consequence of Bayes' theorem, it suggests an important change in viewpoint on generative models: while learning the full $P(a_1,...,a_L)$ from the input MSA ${\cal D}$ is a task of unsupervised learning (sequences are not labeled), learning the factors $P(a_i|a_{i-1},...,a_1)$ becomes a task of supervised learning, with $(a_{i-1},...,a_1)$ being the input (feature) vector, and $a_i$ the output label (in our case a categorical $q$-state label). We can thus build on the full power of supervised learning, which is methodologically more explored than unsupervised learning \cite{bishop2006pattern,hastie2009elements,goodfellow2016deep}.   

In this work, we choose the following parameterization, previously used in the context of statistical mechanics of classical~\cite{wu2019solving} and quantum~\cite{sharir2020deep} systems:
\begin{equation}
  \label{eq:softmax}
  P(a_i|a_{i-1},...,a_1) = \frac { \exp\left\{
      h_i(a_i) + \sum_{j=1}^{i-1}J_{ij}(a_i,a_j) \right\}
  } { z_i(a_{i-1},...,a_1)}\ ,
\end{equation}
with $z_i(a_{i-1},...,a_1) = \sum_{a_i} \exp\{ h_i(a_i) +
\sum_{j=1}^{i-1}J_{ij}(a_i,a_j) \}$ being a normalization factor. In machine learning, this parameterization  is known as soft-max regression, the generalization of logistic regression to multi-class labels~\cite{hastie2009elements}. This choice, as detailed in the section Methods\ref{Methods}, enables a particularly efficient parameter learning by likelihood maximization, and leads to a speedup of 2-3 orders of magnitude over bmDCA, as is reported in Table~\ref{tab:T1}. Because the resulting model is parameterized by a set of fields $h_i(a)$ and couplings $J_{ij}(a,b)$ as in DCA, we dub our method as arDCA.

Besides comparing the performance of this model to bmDCA and DeepSequence, we will also use simple ``fields-only'' models, also known as profile models or independent-site models. In these models, the joint probability of all positions in a sequence factorizes over all positions, $P(a_1,...,a_L)=\prod_{i=1,...,L} f_i(a_i)$, without any conditioning to the sequence context. Using maximum-likelihood inference, each factor $f_i(a_i)$ equals the empirical frequency of amino acid $a_i$ in column $i$ of the input MSA ${\cal D}$.

A few remarks are needed.

Eq.~\eqref{eq:softmax} has striking similarities to standard DCA~\cite{cocco2018inverse}, 
  but also important differences. The two have
  exactly the same number of parameters, but their meaning is quite
  different. While DCA has symmetric couplings
  $J_{ij}(a,b)=J_{ji}(b,a)$, the parameters in Eq.~\eqref{eq:softmax}
  are directed and describe the influence of site $j$ on site $i$ for
  $j<i$ only, {\em i.e.}~only one triangular part of the $J$-matrix is
  filled.
  
The inference in arDCA is very similar to plmDCA~\cite{ekeberg2013improved}, {\em i.e.}~to DCA 
  based on pseudo-likelihood maximization~\cite{balakrishnan2011learning}. 
  In particular, both in arDCA and plmDCA the gradient of the likelihood can be computed exactly from the data, while in bmDCA it has to be estimated via Monte Carlo Markov Chain (MCMC), which requires the introduction of additional hyperparameters (such as the number of chains, the mixing time, etc.) that can have an important impact on the quality of the inference, see~\cite{decelle2021equilibrium} for a recent detailed study.

    In plmDCA each $a_i$ is, however, 
  conditioned to {\em all} other $a_j$ in the sequence, and not only by partial 
  sequences. The resulting directed couplings are usually symmetrized akin to 
  standard Potts models. On the contrary, the $J_{ij}(a,b)$ that appear in arDCA
  cannot be interpreted as ``direct couplings'' in the DCA sense, cf.~below for 
  details on arDCA-based contact prediction. However, plmDCA has limited capacities as a generative model \cite{figliuzzi2018pairwise}: symmetrization moves parameters away from their maximum-likelihood value, probably causing a loss in model accuracy. 
  No such symmetrization is needed for arDCA.

arDCA, contrary to all other DCA methods, allows for calculating the
  probabilities of single sequences. In bmDCA, we can only determine sequence 
  weights, but the normalizing factor, {\em i.e.}~the partition function, remains 
  inaccessible for exact calculations; expensive thermodynamic integration via MCMC 
  sampling is needed to estimate it. The
  conditional probabilities in arDCA are individually normalized; instead of
  summing over $q^L$ sequences we need to sum $L$-times over the $q$
  states of individual amino acids. This may turn out as a major
  advantage when the same sequence in different models shall be
  compared, as in homology detection and protein
  family assignment~\cite{soding2005protein,eddy2009new}, cf.~the example given below.  

 The ansatz in Eq.~\eqref{eq:softmax} can be generalized to more complicated relations. We have tested a two-layer architecture, but did not observe advantages over the simple soft-max regression, as will be discussed at the end of the paper.  

Thanks, in particular, to the possibility of calculating the gradient exactly, arDCA models can be inferred much more efficiently than bmDCA models. Typical inference times are given in Table~\ref{tab:T1} for five representative families, and show a speedup of about 2-3 orders of magnitude with respect to the bmDCA implementation of~\cite{barrat2020sparse}, both running on a single Intel Xeon E5-2620 v4 2.10GHz CPU. We also tested the Mi3 package~\cite{haldane2021mi3}, which is able to learn similar bmDCA models in a time of about 60 minutes for the PF00014 family and 900 minutes for the PF00595 family, while running on two TITAN RTX GPUs, thus remaining much more computationally demanding than arDCA.

\subsection*{The positional order matters}

Eq.~\eqref{eq:ar} is valid for any order of the
positions, {\em i.e.} for any permutation of the natural
positional order in the amino-acid sequences. This is no longer
true, when we parameterize the $P(a_i|a_{i-1},...,a_1)$ according 
to Eq.~\eqref{eq:softmax}. Different orders may give different
results. In the Supplementary Note 1 we show that the likelihood depends on the
order, and that we can optimize over orders. We also find that the
best orders are correlated to the entropic order, where we select first the least entropic, {\em i.e.} most conserved,
variables, progressing successively towards the most variable positions of highest entropy. The site entropy $s_i=-\sum_a f_i(a) \log f_i(a)$ can be directly calculated from the empirical amino-acid frequencies $f_i(a)$ of all amino acids $a$ in site $i$.

Because the optimization over the possible $L!$ site orderings is very time consuming, we use the entropic order as a practical heuristic choice. In all our tests, described in the next sections, the entropic order does not perform significantly worse than the best optimized order we found.

A close-to-entropic order is also attractive from the point of view of interpretation. The most conserved sites come first. If the amino acid on those sites is the most frequent one, basically no information is transmitted further. If, however, a sub-optimal amino acid is found in a conserved position, this has to be compensated by other mutations, {\em i.e.} necessarily by more variable (more entropic) positions. Also the fact that variable positions come last, and are 
modeled as depending on all other amino acids, is well interpretable: 
these positions, even if highly variable, are not necessarily unconstrained, 
but they can be used to finely tune the sequence to any sub-optimal choices 
done in earlier positions. 
 
\begin{table*}[!htb]
    \centering
    \footnotesize
    \begin{tabular}{|l|c|c||c|c|c|c||c|c|c|c||c|c||c|c|}
    \hline  
    &&& $C_{ij}$ ent. & $C_{ij}$ dir. & $C_{ij}$ & $C_{ij}$
    & $C_{ijk}$ ent. & $C_{ijk}$ dir. & $C_{ijk}$ & $C_{ijk}$
    & entropy & entropy & t/min & t/min\\
    & $L$ & $M$ & arDCA & arDCA & bmDCA & DeepSeq
    & arDCA & arDCA & bmDCA & DeepSeq
    & arDCA & bmDCA & arDCA & bmDCA \\
    \hline
    PF00014	& 53 & 13600 & {\bf 0.97}	& 0.96	& 0.95 & 0.81	& {\bf 0.84}	& 0.82	& 0.83 & 0.80	& 1.2	& 1.5 &  {\bf 1} & 204 \\
    PF00076	& 70 & 137605 & {\bf 0.97}	& {\bf 0.97} & {\bf 0.97} & 0.84 & 0.78	& 0.76	& {\bf 0.85} & 0.77 & 1.6	& 1.7 & {\bf 19} & 2088 \\
    PF00595	& 80 & 36690 & 0.96	& 0.95	& {\bf 0.97} & 0.93	& 0.87	& 0.87	& {\bf 0.92} & 0.65 & 1.2	& 1.5 & {\bf 8} & 4003 \\
    PF00072	& 112 & 823798 & {\bf 0.96}	& {\bf 0.96}	& 0.93 & 0.95 & 0.89	& 0.88	& 0.88 & {\bf 0.92}	& 1.4	& 1.8 & {\bf 9} & 1489 \\
    PF13354	& 202 & 7515 & {\bf 0.97}	& 0.96	& 0.95 & 0.95 & {\bf 0.93}	& 0.91	& 0.92 & 0.92 & 0.9	& 1.2 & {\bf 10} & 3905 \\
    \hline
    \end{tabular}
    \caption{The table summarizes the data used (protein families, sequence 
    lengths $L$ and numbers $M$, together with the Pearson correlations between 
    empirical and model-generated connected correlations $C_{ij}$ and $C_{ijk}$,
    for bmDCA, for arDCA using entropic or direct positional orders, and for DeepSequence. The 
    entropies/site and computational running times
    for model learning (on a single Intel Xeon  E5-2620 v4 2.10GHz CPU) are also
    provided for arDCA and bmDCA. 
    Best values for each measure are evidenced. 
    Similar results for the 32 protein families with deep-mutational
    scanning data are given in the Supplementary Table~2.}
    \label{tab:T1}
\end{table*}

For this reason, all coming tests are done using increasing entropic order, {\em i.e.} with sites ordered before model learning by increasing empirical $s_i$ values. Supplementary Figures~1-3 shows a comparison with alternative orderings, 
such as the direct one (from 1 to $L$), several random ones, and the optimized one, cf. also Table~\ref{tab:T1} for some results.

\subsection*{arDCA provides accurate generative models}

To check the generative property of arDCA , we compare it with bmDCA \cite{figliuzzi2018pairwise}, 
{\em i.e.} the most accurate generative version of DCA obtained via Boltzmann 
machine learning. bmDCA was previously shown to be generative not only 
in a statistical sense, but also in a biological one: sequences generated by 
bmDCA were shown to be statistically indistinguishable from natural ones, 
and most importantly, functional {\em in vivo} for the case of chorismate 
mutase enzymes \cite{russ2020evolution}. We also compare the generative property of arDCA with DeepSequence \cite{riesselman2018deep,mcgee2020generative} as a prominent representative of deep generative models.

To this aim, we compare the statistical properties of natural sequences with those of independently and identically distributed (i.i.d.) samples drawn from the different generative models $P(a_1,...,a_L)$. At this point, another important advantage of arDCA comes into play: while generating i.i.d. samples from, {\em e.g.}, a Potts model requires MCMC simulations, which in some cases may have very long
decorrelation times and thus become tricky and computationally expensive \cite{barrat2020sparse,decelle2021equilibrium} (cf.~also Supplementary Note~2 and Supplementary Figure~4),
drawing a sequence from the arDCA model $P(a_1,...,a_L)$ is very simple and does not require any additional parameter. The factorized expression Eq.~\eqref{eq:ar} allows for sampling amino acids position by position, following the chosen positional order, cf.~the detailed description in Supplementary Note~2.

\begin{figure*}[!htb]
	\begin{center}
		\includegraphics[keepaspectratio,width=\textwidth]{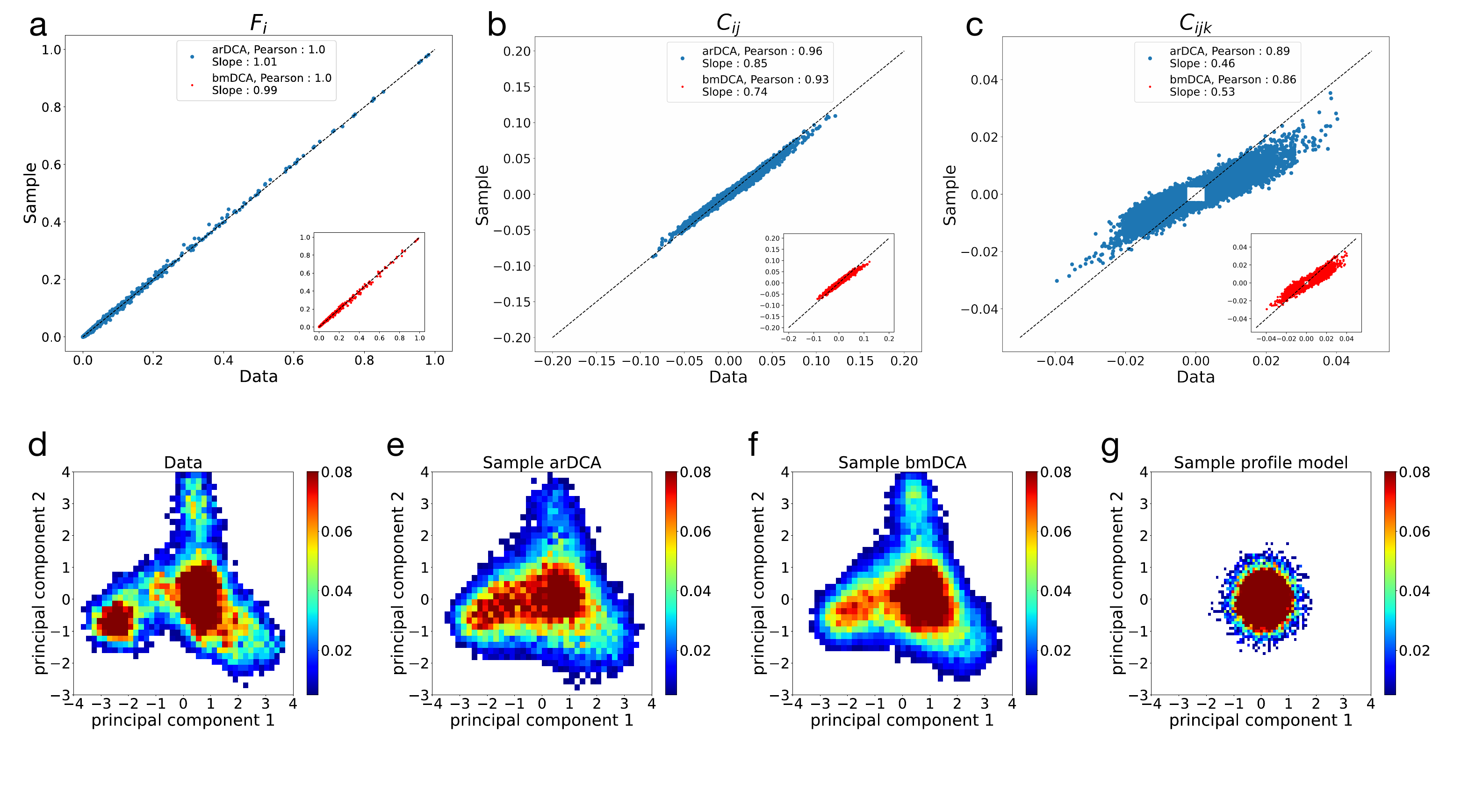}
	\end{center}
	\vspace{-10mm}
	\caption{ Generative properties of arDCA for PF00072: Panels a-c compare the single-site frequencies $f_i(a)$ and two- and three-site connected correlations $C_{ij}(a,b)$ and $C_{ijk}(a,b,c)$ found in the sequence data and samples from models, for arDCA (blue) and bmDCA (red). Panels d-g show different samples projected onto the first two principal components of the natural data. Datasets are the natural MSA (d) and samples from arDCA (e), bmDCA (f) and a profile model (g). Results for other protein families are shown in the Supplementary Figures~5-6.}
        \label{fig:generative}
\end{figure*}

Figures~\ref{fig:generative}a-c show the comparison of the one-point
amino-acid frequencies $f_i(a)$, and the connected two- and
three-point correlations
\begin{eqnarray}
  \label{eq:corrs}
  C_{ij}(a,b) & = & f_{ij}(a,b) - f_{i}(a) f_{j}(b) \ , \\
  C_{ijk}(a,b,c) & = & f_{ijk}(a,b,c) - f_{ij}(a,b) f_{k}(c) -
                       f_{ik}(a,c) f_{j}(b)  \nonumber \\
                   &&    -f_{jk}(b,c) f_{i}(a) + 2 f_{i}(a) f_{j}(b) f_{k}(c) \ ,
  \nonumber
\end{eqnarray}
of the data with those estimated from a sample of the arDCA
model. Results are shown for the response-regulator Pfam family PF00072 \cite{el2019pfam}. 
Other proteins are shown in Table~\ref{tab:T1} and Supplementary Note~3, Supplementary Figures~5-6. We 
find that, for these observables, the empirical and model averages coincide very well, equally well or even slightly better than for the bmDCA case. In particular for the one- and two-point quantities this is quite surprising: while bmDCA fits them explicitly, {\em i.e.} any deviation is due to imperfect fitting of the model, arDCA does not fit them
explicitly, and nevertheless obtains higher precision. 

In Table~\ref{tab:T1}, we also report the results for sequences sampled from DeepSequence \cite{riesselman2018deep}. While its original implementation aims at scoring individual mutations, cf.~Section~\ref{sec:dms}Predicting mutational effects via in-silico deep mutational scanning, we apply the modification of Ref.~\cite{mcgee2020generative} allowing for sequence sampling. We observe that for most families, the two- and three-point correlations of the natural data are significantly less well reproduced by DeepSequence than by both DCA implementations, confirming the original findings of \cite{mcgee2020generative}. Only in the largest family, PF00072 with more than 800,000 sequences, DeepSequence reaches comparable or, in the case of the three-point correlations, even superior performance.

A second test of the generative property of arDCA is given by Figures~\ref{fig:generative}d-g. Panel d
shows the natural sequences projected onto their first two principal
components (PC). The other three panels show generated data projected onto
the same two PCs of the natural data. We see that
both arDCA and bmDCA reproduce quite well the clustered structure of the
response-regulator sequences (both show a slightly broader
distribution than the natural data, probably due to the regularized
inference of the statistical models). On the contrary, sequences
generated by a profile model $P_{\rm prof}(a_1,...,a_L)=\prod_i
f_i(a_i)$ assuming independent sites, do not show any clustered structure: the projections are
concentrated around the origin in PC space. This indicates that their
variability is almost unrelated to the first two principal components
of the natural sequences.

From these observations, we conclude that arDCA provides 
excellent generative models, of at least the same accuracy of bmDCA. This 
suggests fascinating perspectives in terms of data-guided statistical sequence
design: if sequences generated from bmDCA models are functional, also
arDCA-sampled sequences should be functional. But this is obtained at much
lower computational cost, cf.~Table~\ref{tab:T1} and without the need to check for convergence of 
MCMC, which makes the method scalable to much bigger proteins.

\subsection*{Predicting mutational effects via {\em in-silico} deep
  mutational scanning} 
  \label{sec:dms}

\begin{figure*}[!htb]
	\begin{center}
		\includegraphics[keepaspectratio,width=0.95\textwidth]{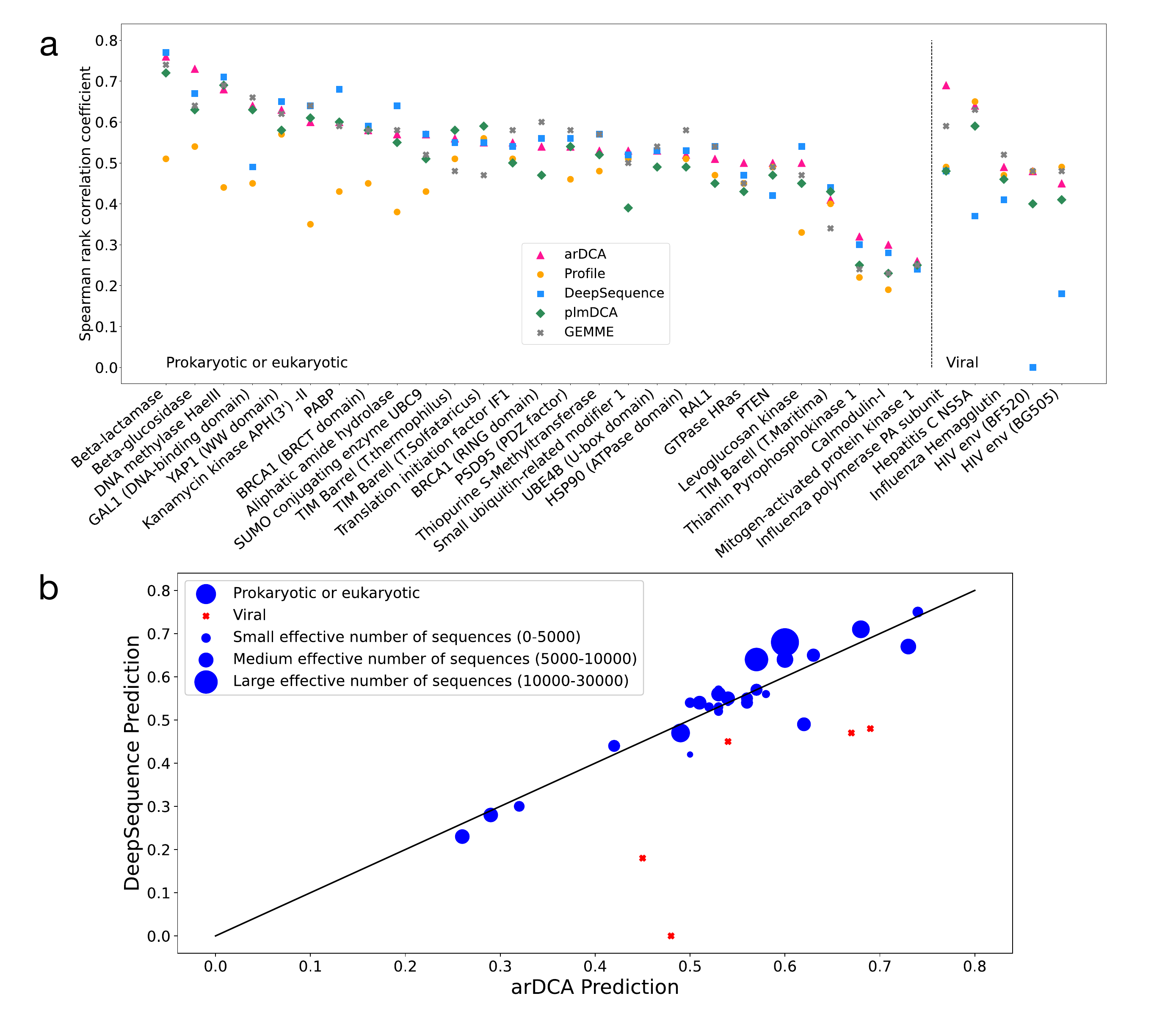}
	\end{center}
	\vspace{-10mm}
	\caption{Prediction of mutational effects by arDCA: Panel a shows the Spearman rank correlation between results of 32 deep-mutational scanning experiments and various computational predictions. We compare arDCA with profile models, plmDCA (aka evMutation \cite{hopf2017mutation}), DeepSequence \cite{riesselman2018deep}, and GEMME \cite{laine2019gemme}, which currently are considered the state of the art. Detailed information about the datasets and the generative properties of arDCA on these datasets are provided in the Supplementary Note~4. Panel b shows a more detailed comparison between arDCA and DeepSequence, the symbol size is proportional to the sequence number in the training MSA for prokaryotic and eukaryotic datasets (blue dots). Viral datasets are indicated by red squares. }
        \label{fig:dms}
\end{figure*}

The probability of a sequence is a measure of its goodness. For high-dimensional
probability distributions, it is generally convenient to work with log-probabilities. Using inspiration
from statistical physics, we introduce a statistical energy
\begin{equation}
  \label{eq:E}
  E(a_1,...,a_L) = - \log P(a_1,...,a_L) \ ,
\end{equation}
as the negative log-probability. We thus expect functional sequences
to have very low statistical energies, while unrelated sequences show 
high energies. In this sense, statistical energy can be seen as a
proxy of (negative) fitness. Note that in the case of arDCA, the statistical
energy is not a simple sum over the model parameters as in DCA, but
contains also the logarithms of the local partition functions
$z_i(a_{i-1},...,a_1)$, cf.~Eq.~\eqref{eq:softmax}.

Now, we can easily compare two sequences differing by one or few
mutations. For a single mutation ${a_i\to b_i}$, where amino
acid $a_i$ in position $i$ is substituted with amino acid $b_i$, we
can determine the statistical-energy difference
\begin{equation}
  \label{eq:delE}
  \Delta E(a_i \to b_i) = - \log
  \frac{P(a_1,...,a_{i-1},b_i,a_{i+1},....,a_L)}{
  P(a_1,...,a_{i-1},a_i,a_{i+1},....,a_L)}
  \ .
\end{equation}
If negative, the mutant sequence has lower statistical energy; the
mutation $a_i\to b_i$ is thus predicted to be beneficial. On the
contrary, a positive $\Delta E$ predicts a deleterious mutation. Note
that, even if not explicitly stated on the left-hand side of
Eq.~\eqref{eq:delE}, the mutational score $\Delta E(a_i \to b_i)$
depends on the whole sequence background $(a_1,
...,a_{i-1},a_{i+1},....,a_L)$ it appears in, {\em i.e.}~on all other
amino acids $a_j$ in all positions $j\neq i$. 

It is now easy to perform an {\em in-silico} deep mutational scan,
{\em i.e.} to determine all mutational scores $\Delta E(a_i \to b_i)$
for all positions $i=1,...,L$ and all target amino acids $b_i$ relative 
to some reference sequence. In Figure~\ref{fig:dms}a, we compare our
predictions with experimental data over more than 30 distinct
experiments and wildtype proteins, and with state-of-the art
mutational-effect predictors. These contain in particular the
predictions using plmDCA (aka evMutation \cite{hopf2017mutation}), variational autoencoders (DeepSequence \cite{riesselman2018deep}),
evolutionary distances between wildtype and the closest homologs
showing the considered mutation (GEMME \cite{laine2019gemme}) -- all of these methods take,
in technically different ways, the context dependence of mutations
into account. We also compare it to the context-independent prediction
using the above-mentioned profile models.

It can be seen that the context-dependent predictors outperform systematically the context-independent predictor, in particular for large MSA in prokaryotic and eukaryotic proteins. The four context-dependent models perform in a very similar way. There is a little but systematic disadvantage for plmDCA, which was the first published predictor of the ones considered here. 

The situation is different in the typically smaller and less diverged viral protein families. In this case, DeepSequence, which relies on data-intensive deep learning, becomes unstable. It becomes also harder to outperform profile models, {\em e.g.} plmDCA does not achieve this. arDCA perform similarly or, in one out of four cases, substantially better than the profile model.

To go into more detail, we have compared more quantiatively the predictions of arDCA and DeepSequence, currently considered as the state-of-the-art mutational predictor. In Figure~\ref{fig:dms}b, we plot the performance of the two predictors against each other, with the symbol size being proportional to the number of sequences in the training MSA of natural homologs. Almost all dots are close to the diagonal (apart from few viral datasets), with 15/32 datasets having a better arDCA prediction, and 17/32 giving an advantage to DeepSequence. The figure also shows that arDCA tends to perform better on smaller datasets, while DeepSequence takes over on larger datasets. In Suppelmentary Figure~7, we have also measured the correlations between the two predictors. Across all prokaryotic and eukaryotic datasets, the two show high correlations in the range of 82\% -- 95\%. These values are larger than the correlations between predictions and experimental results, which are in the range of 50\% -- 60\% for most families. This observation illustrates that both predictors extract a highly similar signal from the original MSA, but this signal may be quite different from the experimentally measured phenotype. Many experiments actually provide only rough proxies for protein fitness, like {\em e.g.} protein stability or ligand-binding affinity. To what extent such variable underlying phenotypes can be predicted by unsupervised learning based on homologous MSA thus remains an open question.

We thus conclude that arDCA permits a fast and accurate prediction of mutational effects, in line with some of the state-of-the-art predictors. It systematically outperforms profile models and plmDCA, and is more stable than DeepSequence in the case of limited datasets. This observation, together with the better computational efficiency of arDCA, suggests that DeepSequence should be used for predicting mutational effects for individual proteins represented by very large homologous MSA, while arDCA is the method of choice for large-scale studies (many proteins) or small families. GEMME, based on phylogenetic informations, astonishingly performs very similarly to arDCA, even if the information taken into account seems different.

\subsection*{Extracting epistatic couplings and predicting
  residue-residue contacts} 

\begin{figure*}[!htb]
	\begin{center}
		\includegraphics[keepaspectratio,width=\textwidth]{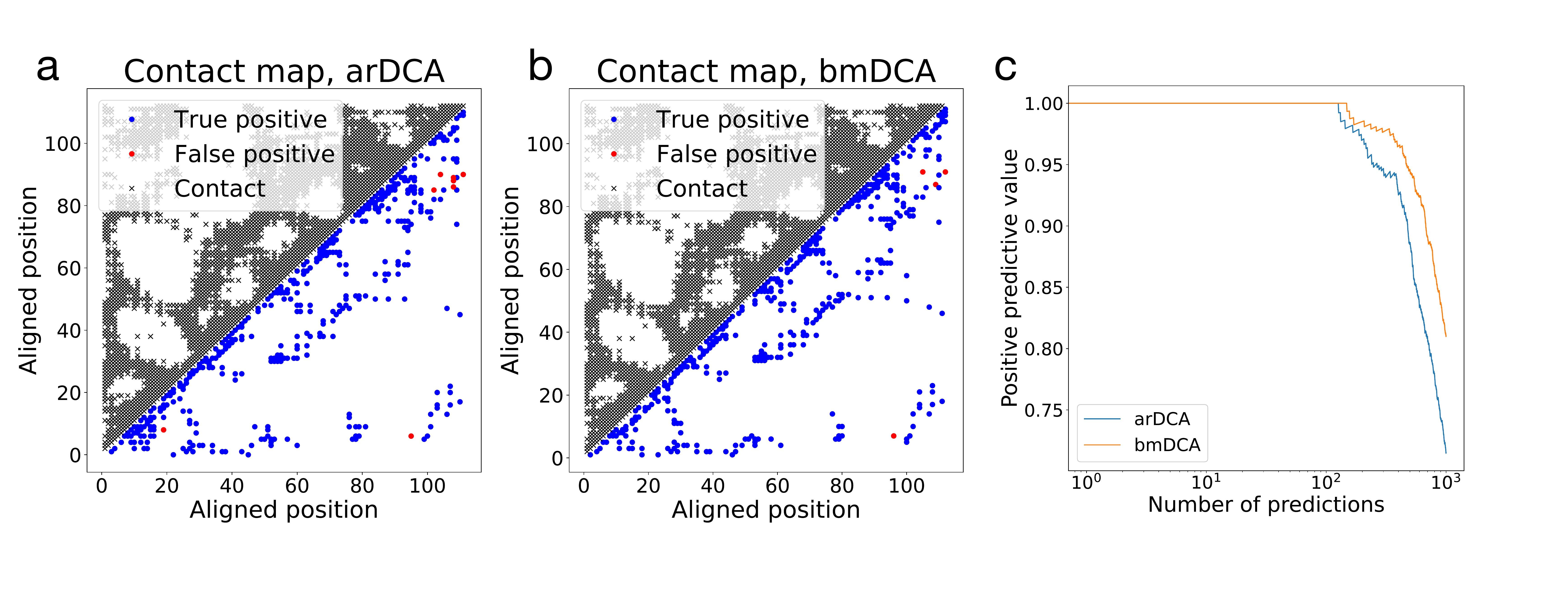}
	\end{center}
	\vspace{-15mm}
	\caption{ Prediction of residue-residue contacts by arDCA as compared to bmDCA. Panels a and b show the true (black, upper triangle) and predicted (lower triangle) contact maps for PF00072, with blue (red) dots indicating true (false) positive predictions. Panel c shows the positive predictive values (PPV, fraction of true positives in the first predictions) as a function of the number of predictions.}
        \label{fig:ppv}
\end{figure*}

The best-known application of DCA is the prediction of residue-residue
contacts via the strongest direct couplings~\cite{morcos2011direct}. As argued before, the arDCA 
parameters are not directly interpretable in terms of direct couplings.
To predict contacts using arDCA, we need to go back to the biological 
interpretation of DCA couplings: they represent epistatic couplings 
between pairs of mutations~\cite{starr2016epistasis}. For a double mutation $a_i\to b_i, a_j\to b_j$, 
epistasis is defined by comparing the effect of the double mutation with 
the sum of the effects of the single mutations, when introduced individually 
into the wildtype background:
\begin{eqnarray}
  \label{eq:epistasis}
  \Delta \Delta E(b_i, b_j) &=& \Delta E(a_i\to b_i, a_j\to b_j) \\
  && - \Delta E(a_i\to b_i) - \Delta E(a_j\to b_j)\ , \nonumber
\end{eqnarray}
where the $\Delta E$ in arDCA are defined in analogy to
Eq.~\eqref{eq:delE}. The epistatic effect $\Delta \Delta E(b_i, b_j)$
provides an effective direct coupling between amino acids
$b_i,b_j$ in sites $i,j$. In standard DCA, $\Delta \Delta E(b_i, b_j)$ is
actually given by the direct coupling
$J_{ij}(b_i,b_j)-J_{ij}(b_i,a_j)-J_{ij}(a_i,b_j)+J_{ij}(a_i,a_j)$
between sites $i$ and $j$.  

For contact prediction, we can treat these effective couplings in the 
standard way (compute the Frobenius norm in zero-sum gauge, apply the 
average product correction, cf.~Supplementary Note~5 for details). The results are 
represented in Figure~\ref{fig:ppv} (cf.~also Supplementary Figures~8-10). The contact maps predicted by arDCA 
and bmDCA are very similar, and both capture very well the topological structure of the native contact map. The arDCA method gives in this case a few more false 
positives, resulting in a slightly lower positive predictive value (panel c). However, note that the majority of the false positives for both 
predictors are concentrated in the upper right corner of the contact maps, 
in a region where the largest subfamily of response-regulators domains, characterized by the coexistence with a Trans\_reg\_C DNA-binding domain (PF00486) in the same protein, has a homo-dimerization interface.

One difference should be noted: for arDCA, the definition of
effective couplings via epistatic effects depends on the reference
sequence $(a_1,...,a_L)$, in which the mutations are introduced;
this is not the case in DCA. So, in principle, each sequence might
give a different contact prediction, and accurate contact prediction in
arDCA might require a computationally heavy averaging over a large
ensemble of background sequences. Fortunately, as we have checked, the 
predicted contacts hardly depend on the reference sequence chosen. It is
therefore possible to take any arbitrary reference sequence belonging 
to the homologous family, and determine epistatic couplings relative 
to this single sequence. This observation causes an enormous speedup 
by a factor $M$, with $M$ being the depths of the MSA of natural 
homologs.

The aim of this section was to compare the performance of arDCA in contact prediction, when compared to established methods using exactly the same data, {\em i.e.}~a single MSA of the considered protein family. We have chosen bmDCA in coherence to the rest of the paper, but apart from little quantitative differences, the conclusions remain unchanged when looking to DCA variants based on mean-field or pseudo-likelihood approximations, cf.~Supplementary Figure~9. The recent success of Deep-Learning--based contact prediction has shown that the performance can be substantially improved if coevolution-based contact prediction for thousands of families is combined with supervised learning based on known protein structures, as done by popular methods like RaptorX, DeepMetaPSICOV, AlphaFold or trRosetta \cite{wang2017accurate,greener2019deep,senior2020improved,yang2020improved}. We expect that the performance of arDCA could equally be boosted by supervised learning, but this goes clearly beyond the scope of our work, which concentrates on generative modeling.

\subsection*{Estimating the size of a family's sequence space}

The MSA of natural sequences contains only a tiny fraction of all sequences, which 
would have the functional properties characterizing a protein family under 
consideration, {\em i.e.} which might be found in newly sequenced
species or be reached by natural evolution. Estimating this number ${\cal N}$ 
of possible sequences, or their entropy $S=\log {\cal N}$, is quite
complicated in the context of DCA-type pairwise Potts models. It requires 
advanced sampling techniques \cite{barton2016entropy,tian2017many}.

In arDCA, we can explicitly calculate the sequence probability $P(a_1,...,a_L)$. 
We can therefore estimate the entropy of the corresponding protein family via
\begin{eqnarray}
  S &=& - \sum_{a_1,...,a_L} P(a_1,...,a_L) \log P(a_1,...,a_L) 
  \nonumber \\
  &=& \langle E(a_1,...,a_L) \rangle_P\ ,
\end{eqnarray}
where the second line uses Eq.~\eqref{eq:E}. The ensemble average 
$\langle \cdot \rangle_P$ can be estimated via the empirical average over a 
large sequence sample drawn from $P$. As discussed before, extracting i.i.d. 
samples from arDCA is particularly simple due to their particular factorized 
form.

Results for the protein families studied here are given in Table~\ref{tab:T1}. As an 
example, the entropy density equals $S/L=1.4$ for PF00072. This corresponds 
to ${\cal N}\sim 1.25 \cdot 10^{68}$ sequences. While being an enormous number, 
it constitutes only a tiny fraction of all $q^L \sim 1.23 \cdot 10^{148}$ 
possible sequences of length $L=112$. Interestingly, the entropies estimated 
using bmDCA are systematically higher than those of arDCA. On the one hand, this 
is no surprise: both reproduce accurately the empirical one- and two-residue 
statistics, but bmDCA is a maximum entropy model, which maximizes the entropy given these statistics~\cite{cocco2018inverse}. On 
the other hand, our observation implies that the effective multi-site 
couplings in $E(a_1,...,a_L)$ resulting from the local partition functions
$z_i(a_{i-1},...,a_1)$ lead to a non-trivial entropy reduction.

\section*{Discussion}

We have presented a class of simple autoregressive models,
which provide highly accurate and computationally very 
efficient generative models for protein-sequence families. While being 
of comparable or even superior performance to bmDCA across a number of 
tests including the sequence statistics, the sequence distribution in 
dimensionally reduced principal-component space, the prediction of
mutational effects and residue-residue contacts, arDCA is computationally
much more efficient than bmDCA. The particular factorized form of 
autoregressive models allows for exact likelihood maximization. 

It allows also for the calculation of exact sequence probabilities (instead 
of sequence weights for Potts models). This fact is of great potential 
interest in homology detection using coevolutionary models, which
requires to compare probabilities of the same sequence in distinct 
models corresponding to distinct protein families. To illustrate this idea in a simple, but instructive case, we have identified two subfamilies of the PF00072 protein family of response regulators. The first subfamily is characterized by the existence of a DNA-binding domain of the Trans\_reg\_C protein family (PF00486), the second by a DNA-binding domain of the GerE protein family (PF00196). For each of the two subfamilies, we have extracted randomly 6,000 sequences used to train sub-family specific profile and arDCA models, with $P_1$ being the model for the Trans\_reg\_C and $P_2$ for the GerE sub-family. Using the log-odds ratio $\log\{P_1(seq)/P_2(seq)\}$ to score all remaining sequences from the two subfamilies, the profile-model was able to assign 98.6\% of all sequences to the correct sub-family, and 1.4\% to the wrong one. arDCA has improved this to 99.7\% of correct, and only 0.3\% of incorrect assignments, reducing the grey-zone in sub-family assignment by a factor 3-4. Furthermore, some of the false assignments of the profile model had quite large scores, cf.~the histograms in Supplementary Figure~11, while the false annotations of the arDCA model had scores closer to zero. Therefore, if we consider that a prediction is reliable only if there is no wrong predictions for a  larger log-odds ratio score, then the score of arDCA is 97.5\% while the one of the profile model is only 63.7\%. 

The importance of accurate generative models becomes also visible via 
our results on the size of sequence space (or sequence entropy). For 
the response regulators used as example throughout the paper (and similar 
observations are true for all other protein families we analyzed), we
find that ``only'' about $10^{68}$ out of all possible $10^{148}$ amino-acid 
sequences of the desired length are compatible with the arDCA model, and
thus suspected to have the same functionality and the same 3D structure 
of the proteins collected in the Pfam MSA. This means that a random
amino-acid sequence has a probability of about $10^{-80}$ to be actually
a valid response-regulator sequence. This number is literally astronomically 
small, corresponding to the probability of hitting one particular atom when
selecting randomly in between all atoms in our universe. The importance of a good coevolutionary modeling 
becomes even more evident when considering all proteins being compatible
with the amino-acid conservation patterns in the MSA: the corresponding 
profile model still results in an effective sequence number of $10^{94}$,
{\em i.e.} a factor of $10^{26}$ larger than the sequence space respecting 
also coevolutionary constraints. As was verified in experiments, conservation
provides insufficient information for generating functional proteins, while
taking coevolution into account leads to finite success probabilities.

Reproducing the statistical features of natural sequences does not necessarily guarantee the sampled sequences to be fully functional protein sequences. To enhance our confidence in these sequences, we have performed two tests.

First we have reanalyzed the bmDCA-generated sequences of \cite{russ2020evolution}, which were experimentally tested for their {\em in-vivo} chorismate-mutase  activity. Starting from the same MSA of natural sequences, we have trained an arDCA model and calculated the statistical energies of all non-natural and experimentally tested sequences. As is shown in Supplementary Figure~12, the statistical energies have a Pearson correlation of 97\% wit the bmDCA energies reported in \cite{russ2020evolution}. In both cases functional sequences are restricted to the region of low statistical energies.

 Furthermore, we have used small samples of 10 artificial or natural response-regulator sequences as inputs for trRosetta \cite{yang2020improved}, in a setting which allows for protein-structure prediction based only on the user-provided MSA, {\em i.e.}~no homologous sequences are added by trRosetta, and no structural templates are used. As is shown in Supplementary Figure~13, the predicted structures are very similar to each other, and within a root mean-square deviation of less than 2\AA~from an exemplary PDB structure. The contacts maps extracted from the trRosetta predictions are close to identical.

While these observation do not prove that arDCA-generated sequences are functional or fold into the correct tertiary structure, they are coherent with this conjecture.

Autoregressive models can be easily extended by adding hidden layers in the ansatz for the conditional probabilites $P(a_i|a_{i-1},...,a_1)$, with the aim to increase the expressive power of the overall model.
For the families explored here, we found that the one-layer model Eq.~\eqref{eq:softmax} is already so accurate, that adding more layers only results in similar, but not superior performance, cf.~Supplementary Note~6. However, in longer or more complicated protein families, the larger expressive power of deeper autoregressive models could be helpful.
Ultimately, the generative performance of such extended models should be assessed by testing the functionality of the generated sequences in experiments similar to~\cite{russ2020evolution}.\\

\section*{Methods}
\label{Methods}
\subsection*{Inference of the parameters}
\label{app:1}

We first describe the inference of the parameters via likelihood maximization. 
In a Bayesian setting, with uniform prior (we discuss regularization below),
the optimal parameters are those that maximize the probability of the data,
given as a MSA ${\cal D}=(a_i^m|i=1,...,L; m=1,...,M)$ of $M$ sequences of aligned length $L$:
\begin{align}
\nonumber\{\textbf{J$^*$,h$^*$}\}&=\arg\max_{\{\textbf{J,h}\}} P({\cal D}|\{\textbf{J,h}\})\\
\nonumber&=\arg\max_{\{\textbf{J,h}\}} \log P({\cal D}|\{\textbf{J,h}\})\\
 \nonumber&= \arg\max_{\{\textbf{J,h}\}} \sum_{m = 1}^{M} \log{\prod_{i=1}^{L}{P(a_i^m|a_{i-1}^m,...,a_1^m)}} \\
           &= \arg\max_{\{\textbf{J,h}\}} \sum_{m = 1}^{M}\sum_{i=1}^{L} \log{P(a_i^m|a_{i-1}^m,...,a_1^m)} \ .  \label{eq:maximization}
\end{align}
Each parameter $h_i(a)$ or $J_{ij}(a,b)$ appears in only one conditional probability $P(a_i|a_{i-1},...,a_1)$, and we can thus maximize independently each conditional probability in Eq.~\eqref{eq:maximization}: 
\begin{align*}
\{\textbf{J$_{ij}^*$,h$_i^*$}\} 
&= \arg\max_{\{\textbf{J$_{ij}$,h$_i$}\}}\sum_{m = 1}^{M} \log{P(a_i^m|a_{i-1}^m,...,a_1^m)} \\
&=\arg\max_{\{\textbf{J$_{ij}$,h$_i$}\}}\sum_{m = 1}^{M}  \Bigg[ h_i(a_i^m)+\sum_{j=1}^{i-1} J_{ij}(a_i^m,a_j^m)\\
& \hspace{3.7cm}   -\log{z_i(a_{i-1}^m,...a_1^m)} \Bigg]
\end{align*}
where
\begin{equation}\label{eq:zdef}
z_i(a_{i-1},...a_1) = \sum_{a_i} \exp\left\{ h_i(a_i) +
\sum_{j=1}^{i-1}J_{ij}(a_i,a_j) \right\}
\end{equation}
is the normalization factor of the conditional probability of variable $a_i$.

Differentiating with respect to $h_i(a)$ or to $J_{ij}(a,b)$, with $j=1,...,i-1$,  we get the set of equations: 
\begin{equation}
\begin{split}
&0=\frac{1}{M}\sum_{m = 1}^{M}  \left[\delta_{a,a_i^m}-  \frac{\partial\log z_i(a_{i-1}^m,...a_1^m)}{\partial h_i(a)}\right] \ ,\\
&0=\frac{1}{M}\sum_{m = 1}^{M}  \left[\delta_{a,a_i^m}\delta_{b,a_j^m}-\frac{\partial\log z_i(a_{i-1}^m,...a_1^m)}{\partial J_{ij}(a,b)}\right]  \ ,
\end{split}
\end{equation}
where $\delta_{a,b}$ is the Kronecker symbol. Using Eq.~\eqref{eq:zdef} we find 
\begin{equation}
\begin{split}
 &\frac{\partial\log z_i(a_{i-1}^m,...a_1^m)}{\partial h_i(a)}=P(a_i = a|a_{i-1}^m,...,a_1^m) \ , \\
 & \frac{\partial\log z_i(a_{i-1}^m,...a_1^m)}{\partial J_{ij}(a,b)}  = P(a_i= a|a_{i-1}^m,...,a_1^m)
\delta_{a_j^m,b} \ .
\end{split}
\end{equation}
The set of equations thus reduces to a very simple form: 
\begin{equation}\label{eq_max_AR}
\begin{split}
f_{i}(a)  &= \left\langle P(a_i= a|a_{i-1}^m,...,a_1^m)\right\rangle_{\cal D}\ , \\
f_{ij}(a,b)   &= \left\langle P(a_i= a|a_{i-1}^m,...,a_1^m)\delta_{a_j^m,b}\right\rangle_{\cal D} \ ,
\end{split}
\end{equation}
where 
$\left\langle\bullet\right\rangle_{\cal D} = \frac1M \sum_{m=1}^M \bullet^m$ denotes the empirical data average, and $f_{i}(a)$, $f_{ij}(a,b)$ are the empirical one- and two-point amino-acid frequencies.
Note that for the first variable ($i=1$), which is unconditioned, there is no equation for the couplings, and the equation for the field takes the simple form $f_1(a) = P(a_1=a)$, which is solved by $h_1(a) = \log f_1(a) + \text{const.}$

Unlike the corresponding equations for the Boltzmann learning of a Potts model~\cite{figliuzzi2018pairwise}, there is a mix between probabilities and empirical averages in Eq.~(\ref{eq_max_AR}), and there is no explicit equality between one- and two-point marginals and empirical one and two-point frequencies. This means that the ability to reproduce the empirical one- and two-point frequencies is already a statistical test for the generative properties of the model, and not only for the fitting quality of the current parameter values.

The inference can be done very easily with any algorithm using gradient descent, which updates the fields and couplings proportionally to the difference of the two sides of Eq.~\eqref{eq_max_AR}. We used the Low Storage BFGS method to do the inference.
We also add a $L2$ regularization, with regularization strength of $0.0001$ for the generative tests and $0.01$ for mutational effects and contact prediction. A small regularization leads to better results on generative tests, but a larger regularization is needed for contact prediction or mutational effects. Contact prediction can indeed suffer from too large parameters, and therefore a larger regularization was chosen, coherently with the one used in plmDCA. Note that the gradients are computed exactly at each iteration, as an explicit average over the data, and hence without the need of MCMC sampling. This provides an important advantage over Boltzmann-machine learning.

Finally, in order to partially compensate for the phylogenetic structure of the MSA, which induces correlations among sequences,
each sequence is reweighted by a coefficient $w_{m}$~\cite{cocco2018inverse}:
\begin{equation}
    \{\textbf{J$_{ij}^*$,h$_i^*$}\} = \arg\max_{\{\textbf{J$_{ij}$,h$_i$}\}} \frac 1{M_{\rm eff}} \sum_{m = 1}^{M}  w_{m} \log P(\textbf{a}^m |\{\textbf{J$_{ij}$,h$_i$}\}) \ ,
 \end{equation}
 which leads to the same equations as above with the only modification of the empirical average as
 $\left\langle\bullet\right\rangle_{data} =  \frac 1{M_{\rm eff}}\sum_{m=1}^M w_m\, \bullet^m$.
Typically, $w_m$ is given by the inverse of the number of sequences having least $80\%$ sequence identity with sequence $m$, and $M_{\text{eff}} = \sum_m w_m$ denotes the effective number of independent sequences. The goal is to remove the influence of very closely related sequences. Note however that such reweighting cannot fully capture the hierarchical structure of phylogenetic relations between proteins.

\subsection*{Sampling from the model}
\label{sampling}

Once the model parameters are inferred, a sequence can be iteratively generated by the following procedure: 
\begin{enumerate}
    \item Sample the first residue from $P(a_1)$ 
    \item Sample the second residue from $P(a_2|a_1)$ where $a_1$ is sampled in the previous step.\\
    ...
    \item[$L$.]Sample the last residue from $P(a_L|a_{L-1},a_{L-2},...,a_2,a_1)$
\end{enumerate}
Each step is very fast because there are only $21$ possible values for each probability. 
Both training and sampling are therefore extremely simple and computationally efficient in arDCA.

\begin{acknowledgments}
We thank Indaco Biazzo, Matteo Bisardi, Elodie Laine, Anna-Paola
Muntoni, Edoardo Sarti and Kai Shimagaki for helpful discussions and
assistance with the data.  
We especially thank Francisco McGee and Vincenzo Carnevale for providing generated samples from DeepSequence as in Ref.~\cite{mcgee2020generative}.
Our work was partially funded by the EU
H2020 Research and Innovation Programme MSCA-RISE-2016 under Grant
Agreement No. 734439 InferNet. J.T. is supported by a PhD Fellowship of the i-Bio Initiative from the Idex Sorbonne University Alliance.\\
\end{acknowledgments}

\noindent {\bf Author contributions}: 
A.P., F.Z. and M.W. designed research; J.T., G.U. and A.P. performed research; J.T., G.U., A.P., F.Z. and M.W. analyzed the data; J.T., F.Z. and M.W. wrote the paper.\\

\noindent {\bf Competing interests}:
The authors declare no competing interests.\\

\noindent {\bf Code availability}:
Codes in Python and Julia are available at https://github.com/pagnani/ArDCA.git.\\

\noindent {\bf Data availability}:
Data is available at https://github.com/pagnani/ArDCAData and was elaborated using source data freely downloadable from the Pfam database (http://pfam.xfam.org/) \cite{el2019pfam}, cf.~Supplementary Table 1. The repository contains also sample MSA generated by arDCA. The input data for Figure 3 are provided by the GEMME paper \cite{laine2019gemme}, cf.~also Supplementary Table 2.

\begin{widetext}
\appendix
\beginsupplement

\section{Positional order}

For a family of length $L$, there are $L!$ possible permutations of the sites and therefore $L!$ possible orders. The parameterization of the conditional probabilities of the arDCA model is not invariant under a change of order, thus different orders may give different results. However, an optimization over all the different orders is not computationally feasible. We compared some particular orders: the direct order along the protein chain, the entropic order where the sites are ordered in ascending order according to their local entropy
$ s_i = -\sum_{a = 1}^{q} f_i(a) \log{f_i(a)}$, and the inverse entropic order where $s_i$ is used in descending order. A comparison with $100$ random orders was also made. The quality of the generative properties was found to be highly correlated with the log-likelihood of the optimized model, which can be computed exactly after the parameters are inferred, see Section Methods of the main text. Supplementary Figure~\ref{likelihood_order_fig} shows a comparison of the likelihood and the Pearson's correlation of the two-point statistics for the different orders. The values reported  for the random order is an average over the $100$ different realizations, with one standard deviation given by the vertical bar. While the direct order is compatible with a random order, for all but one family the entropic order has the highest value of the likelihood and maximizes the Pearson correlation.

\begin{figure}
      \includegraphics[scale = 0.55]{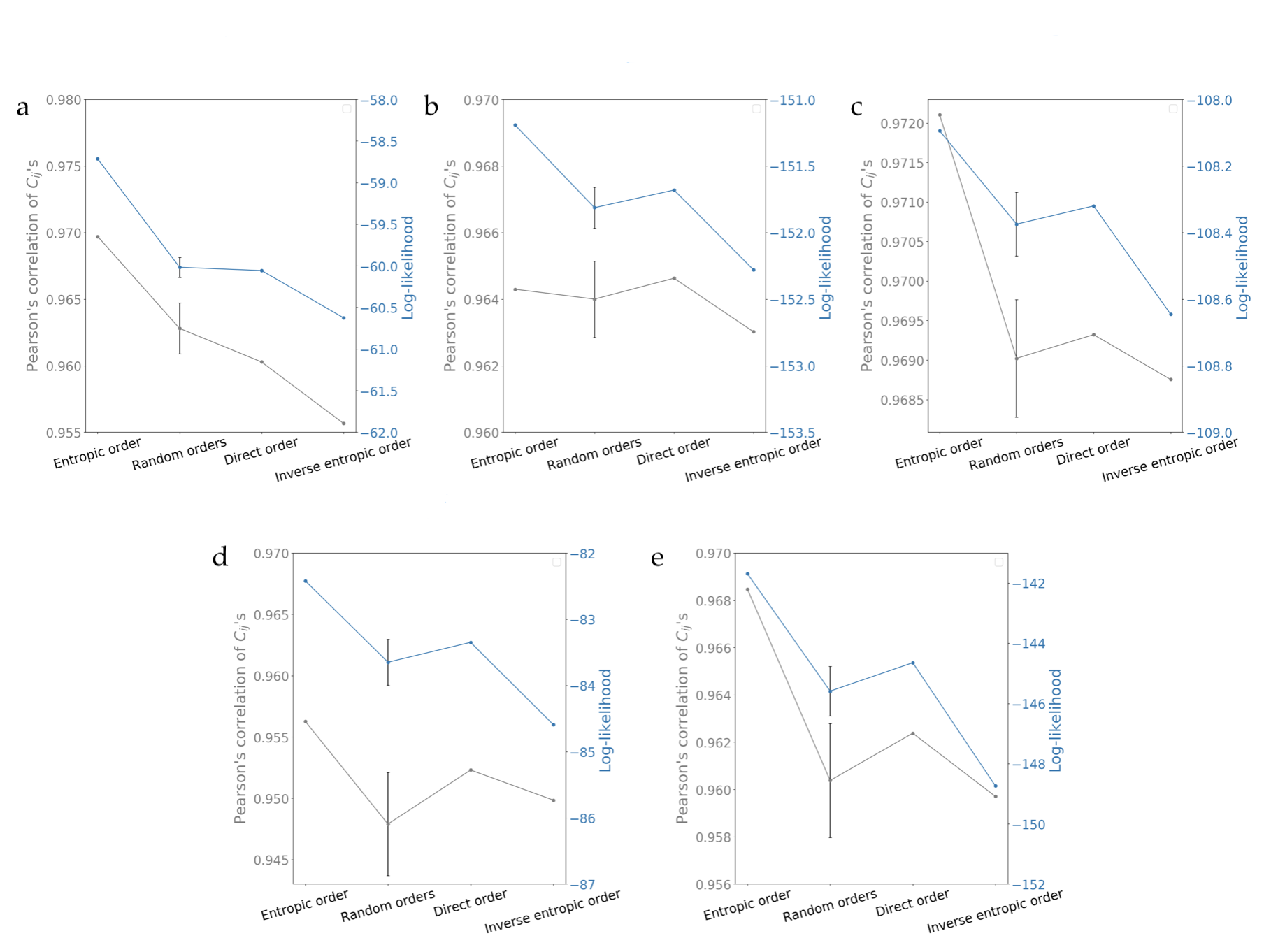}
      \caption{Comparison of the Pearson correlation of two-point connected correlations and the log-likelihood between different orders. The value of the random order is the mean of $100$ random orders with one standard deviation given as a vertical bar. Results are for Pfam families PF00014 (a), PF00072 (b), PF00076 (c), PF0595 (d), and PF13354 (e).}
      \label{likelihood_order_fig}
\end{figure}

In order to check that the entropic order is a good heuristic choice between all possible orders, we designed a greedy procedure to increase the likelihood by doing some permutations between the sites. This procedure tries to find a locally optimal order. The different steps of the procedure are:
\begin{itemize}
    \item Choose a site randomly
    \item Permute this site with all the other sites and compute the likelihood of the new model each time
    \item Choose the permutation that increases the most the likelihood and iterate the procedure
\end{itemize}
Supplementary Figure~\ref{permutations}a shows the evolution of the log-likelihood of the entropic and direct orders of the family PF00014 under permutations. The permuted entropic order saturates quickly to a value of the log-likelihood relatively close to the initial one, indicating that the entropic order is not far from a locally optimal one. The direct order saturates to the same value of the log-likelihood. The plot also shows the evolution of the Kendall-Tau distance between the two orders, defined as the number of pairs in a different order, i.e. for two lists $l1$ and $l2$,
\begin{equation}
    K(l1,l2) = | \{(i,j): i<j, (l1(i)<l1(j) \cap l2(i)>l2(j)) \cup  (l1(i)>l1(j) \cap l2(i)<l2(j)) \} | \ .
\end{equation}
The Kendall-Tau distance gives a measure of the dissimilarity between two lists. Supplementary Figure~\ref{permutations}a shows that the distance between the two orders decreases with increasing permutations. Supplementary Figures~\ref{permutations}b and 
\ref{permutations}c show the value of the local entropy for each site in both orders before (b) and after (c) $60$ permutations. After $60$ permutations, it is clear that the sites with a low local entropy are typically at the beginning, which is coherent with the explanation given in the main text. 

\begin{figure}
      \centering
      \includegraphics[scale = 0.5]{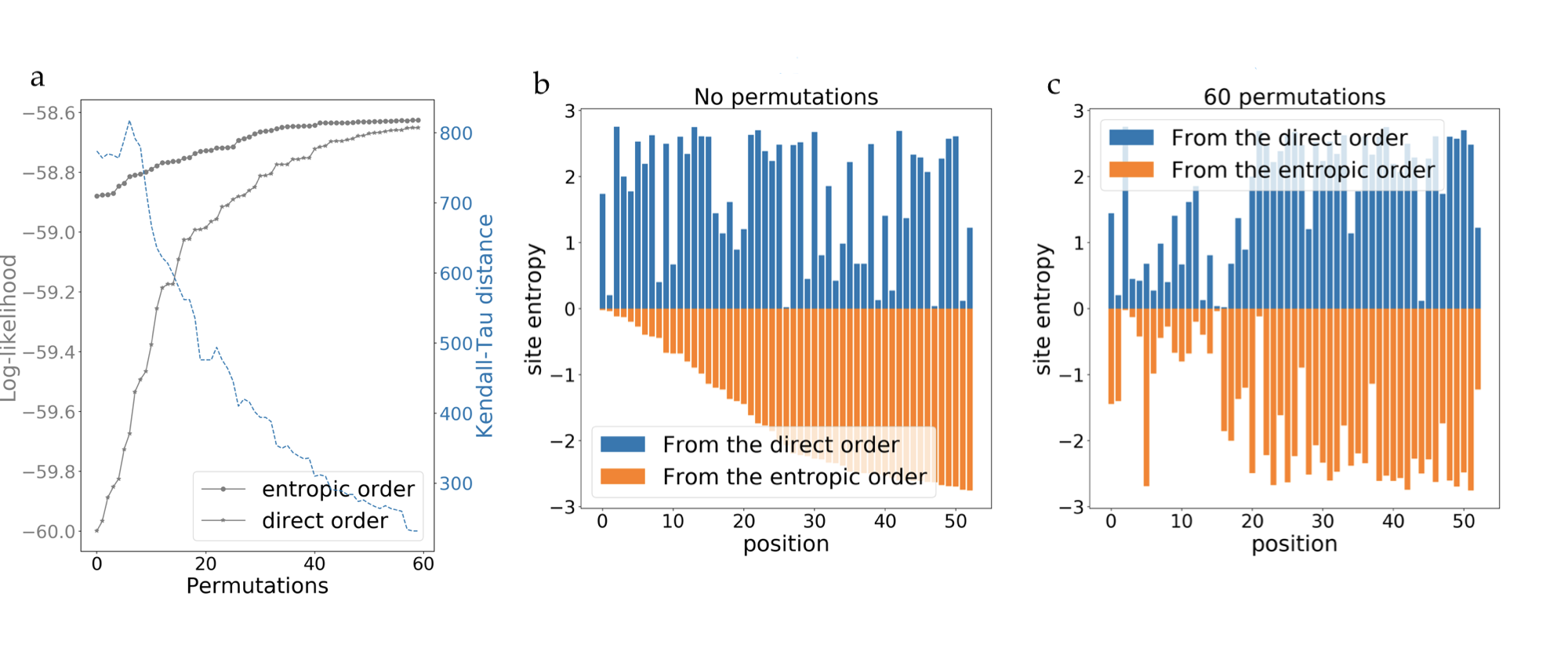}
      \caption{a: Evolution of the log-likelihood and the Kendall-Tau distance of the direct and entropic orders under permutations. b and c: Values of the entropy of each site in the direct and entropic orders with no permutations~(b) and after 60 permutations~(c).}
       \label{permutations}
\end{figure}

Finally, Supplementary Figure~\ref{cij_order} shows the Pearson correlation of two-point connected correlations for the $33$ families used for mutational effects with the entropic and the direct order. Coherently with the previous discussion, the entropic order gives a better result for $30$ over $33$ families.

\begin{figure}[t]
      \centering
      \includegraphics[scale = 0.2]{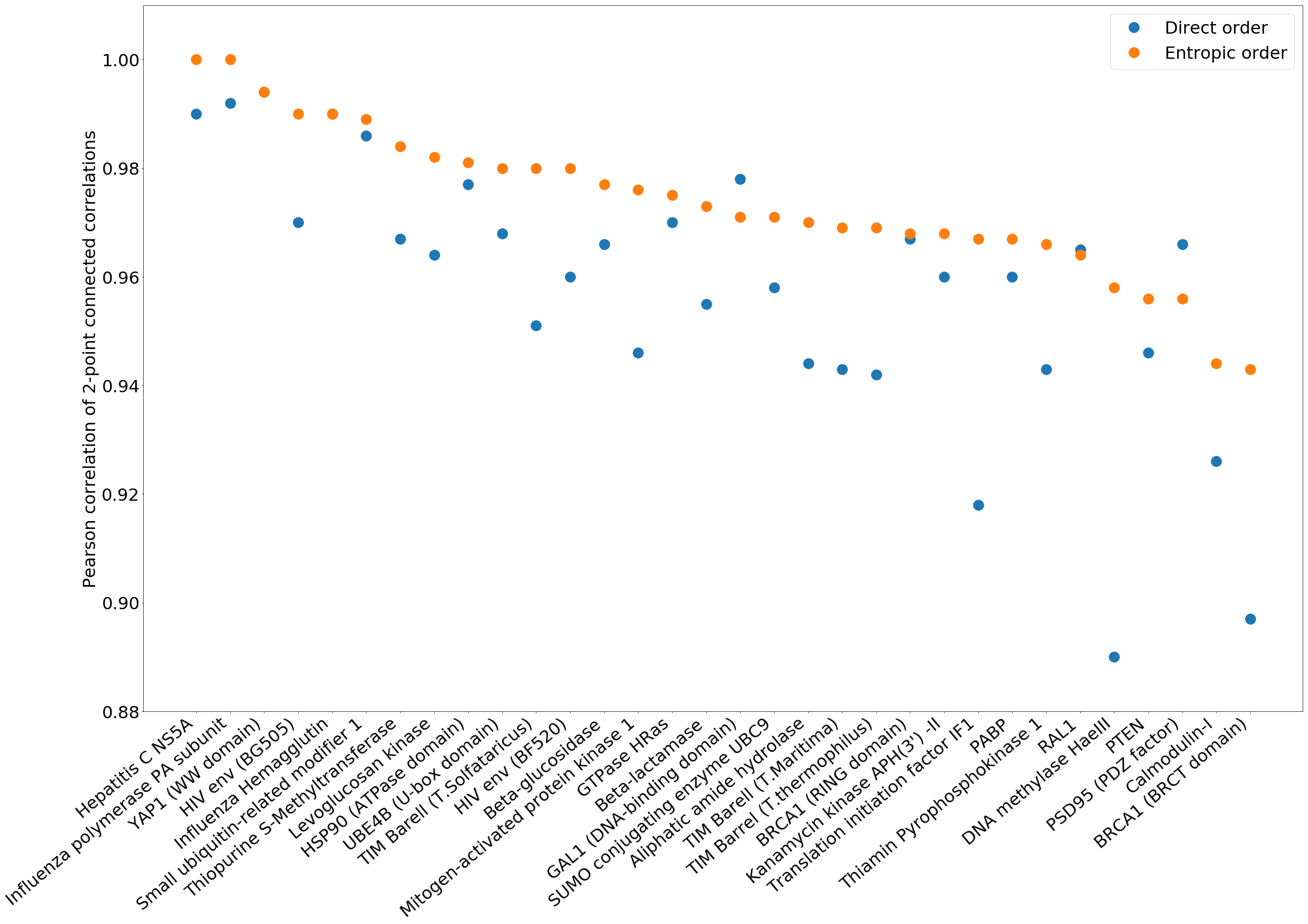}
      \caption{Pearson correlation of the two-point connected correlations for all the 33 families used for mutational effects, for the direct and entropic orders.}
      \label{cij_order}
\end{figure}

\clearpage

\section{Sampling from the model}

To emphasize the advantages of the direct sampling protocol of arDCA, we report here 
a comparison with sampling from the bmDCA model.
In this case, sequences must be obtained via MCMC sampling, but a lot of moves have to be made in order to achieve a proper equilibration in some families~\cite{figliuzzi2018pairwise,barrat2020sparse,decelle2021equilibrium}. Furthermore, several hyperparameters have to be set, such as the number of MCMC independent chains, the total length and the number of samples produced by each chain, etc. As discussed in detail in~\cite{decelle2021equilibrium}, these hyperparameters heavily affect the quality of the training and sampling in bmDCA. If the mixing time of MCMC is short enough, then it is possible to train and sample the bmDCA model in equilibrium, which leads to stable and reproducible results. If the mixing time is too long, however, this becomes impossible and one is forced to train the machine out-of-equilibrium (e.g. via CD or PCD), which leads to an unstable resampling displaying a maximal quality at some given sampling time that depends on the training history~\cite{decelle2021equilibrium}.

\begin{figure}[t]
    \centering
    \includegraphics[width=.8\textwidth]{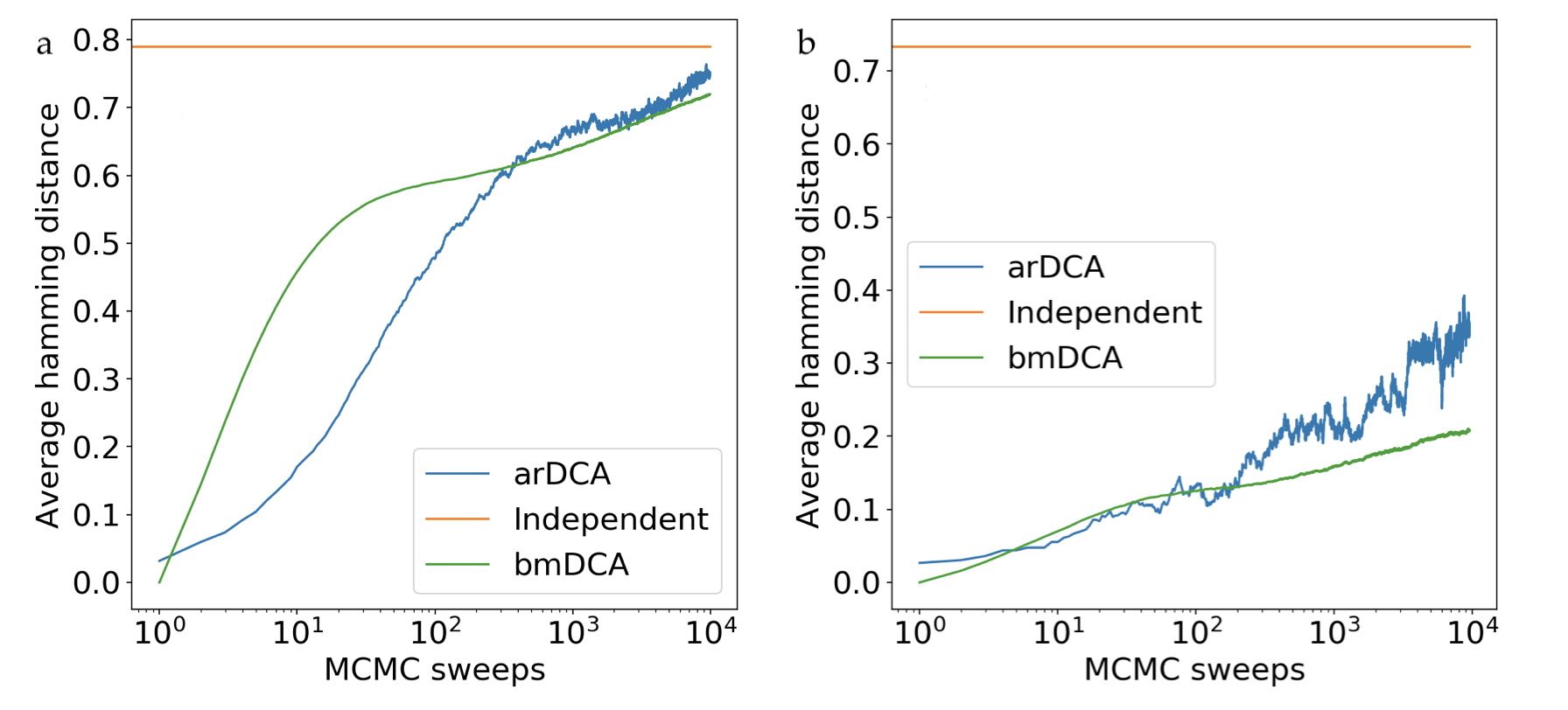}
    \caption{Averaged Hamming distance between a sequence and its time evolution after MCMC sweeps, in bmDCA and in arDCA. The average is made with respect to $100$ (panel a, PF13354 family) or $5$ (panel b, PF18589 family) initially equilibrated sequences. The horizontal line is the average Hamming distance between two independent equilibrium sequences, which by the direct arDCA-sampling procedure can be achieved in the equivalent of a single sweep (each positions sampled once).
    }
    \label{MCMC}
\end{figure}

As an illustration of these effects, we consider a beta-lactamase family (Pfam family PF13354), which is particularly hard to sample via MCMC; bmDCA models have then been trained via PCD~\cite{barrat2020sparse}.
In order to directly compare with MCMC sampling of bmDCA, we sampled sequences from our arDCA model, using a Metropolis-Hasting procedure. We propose a random change of a residue, and we accept the move with a probability that depends on the ratio of the probabilities of the new and old sequences. This is of course a very inefficient way of sampling from the arDCA model, but it allows for a direct comparison with the MCMC dynamics of a bmDCA.
The Hamming distance between an initial equilibrium sequence (obtained via the sequential procedure described above, which thus guarantees equilibration) and its time evolution after MCMC sweeps was computed. This time-dependent Hamming distance, averaged over $100$ initial sequences, is reported in Supplementary Figure~\ref{MCMC}a. Its shape is very similar to that obtained by MCMC sampling of bmDCA~\cite{barrat2020sparse}, also reported in the same figure (note that in this case the initial sequence is not fully equilibrated).
It grows very slowly with time, and only at very long times it saturates to the equilibrium Hamming distance between two independently sampled sequences. The time it takes to reach this plateau gives an estimation of the number of MCMC sweeps needed to obtain an equilibrium sample. Supplementary Figure~\ref{MCMC}a shows that the equilibration takes at least $10^4$ MCMC sweeps. On the other hand, the sequential procedure described above, which is only possible for arDCA models, allows one to sample almost instantaneously, thus completely bypassing the long time scale associated to MCMC. 

We repeated the same study for the obesity receptor family (ObR\_IG, Pfam family PF18589) and for the Leptin family (Pfam family PF02024), where the number of available sequences is more limited and bmDCA has shown some convergence problems. We find that the mixing time of MCMC is even larger in that case, see Supplementary Figure~\ref{MCMC}b for PF18589. As a result, the bmDCA training is strongly out-of-equilibrium. While the Pearson coefficient of the $C_{ij}$ reaches $0.99$ during training, a resampling of the same model leads to very poor results ($\sim0.62$ for PF18589 and $\sim0.43$ for PF02024). We conclude that bmDCA, at least in our simplest scheme, is not reliable. On the contrary, arDCA provides reliable results for both families.

We note that these observations have interesting implication for the problem of sampling in disordered systems with slow dynamics, as already noted in~\cite{wu2019solving}.

\section{Results for other families}
\subsection{Pfam Datasets}
\subsubsection{Description
}
We describe the properties of the five Pfam families used to test the generative properties and structure prediction of the arDCA model: PF00014, PF00072, PF00076, PF00595, PF13354. MSA are downloaded from Pfam (http://pfam.xfam.org/) and sequences with more than 6 consecutive gaps are removed. The value of $M_{\rm eff}$ defined in Section Methods of the main text gives the effective number of sequences, obtained
by a proper reweighing of very similar sequences. 

\begin{table}[h]
\begin{tabular}{>{\centering}p{0.15\textwidth}p{0.15\textwidth}p{0.15\textwidth}p{0.15\textwidth}p{0.15\textwidth}p{0.15\textwidth}}
\hline
 Pfam identifier& PF00014 & PF00072 & PF00076 & PF00595 & PF13354\\\hline
 Protein domain & Kunitz domain& Response regulator & RNA recognition & PDZ domain & Beta-lactamase\\
 & & receiver domain & motif &  & \\
 \hline
  $L$ &53& 112 & 70 & 82 & 202\\\hline
   $M$ &13600& 823798 & 137605  & 36690  & 7515\\\hline
   $M_{\rm eff}$ &4364 & 229585  & 27785  & 3961  & 7454\\\hline
\end{tabular}
\caption{Properties of the Pfam families}
\end{table}

\subsubsection{Principal component analysis}

Supplementary Figure~\ref{PCA} shows the projection of natural sequences (first column), sequences sampled from arDCA (second column), bmDCA (third column) and the profile model (last column) in a two-dimensional space, constructed by performing principal component analysis on the natural sequences. Each bin in the figure has a color related to its total weight, defined by resampled the sequences using the weights defined in Section Methods of the main text.

\begin{figure}
\centering
\includegraphics[scale=0.6]{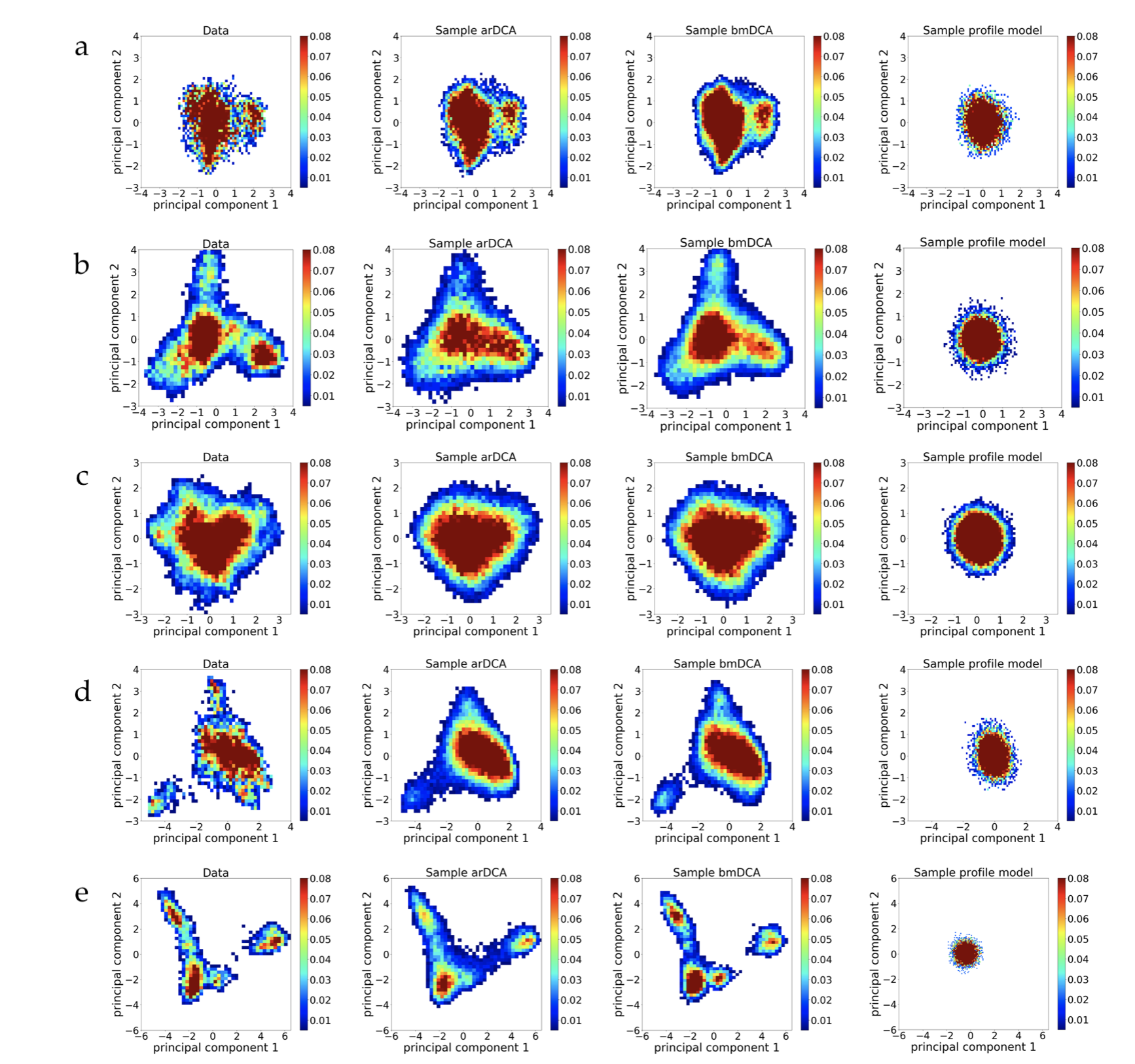}
\caption{Projections of sequences on the principal components obtained from natural sequences, for the Pfam families PF00014 (a), PF00072 (b), PF00076 (c), PF0595 (d), and PF13354 (e).}
\label{PCA}
\end{figure}

\subsubsection{Frequencies}
Supplementary Figure~\ref{freq} shows how well the model is able to reproduce the empirical frequencies obtained from the data. The one-point frequencies (left), two-point (center) and three-point (right) connected correlations are shown, both from the arDCA model (blue) and the bmDCA (red). Note that for the three-point connected correlations, the correlations that have an empirical value smaller than $0.003$ are removed, because they are not meaningful given the limited number of sequences in the dataset. 

\begin{figure}
\centering
\includegraphics[scale=0.33]{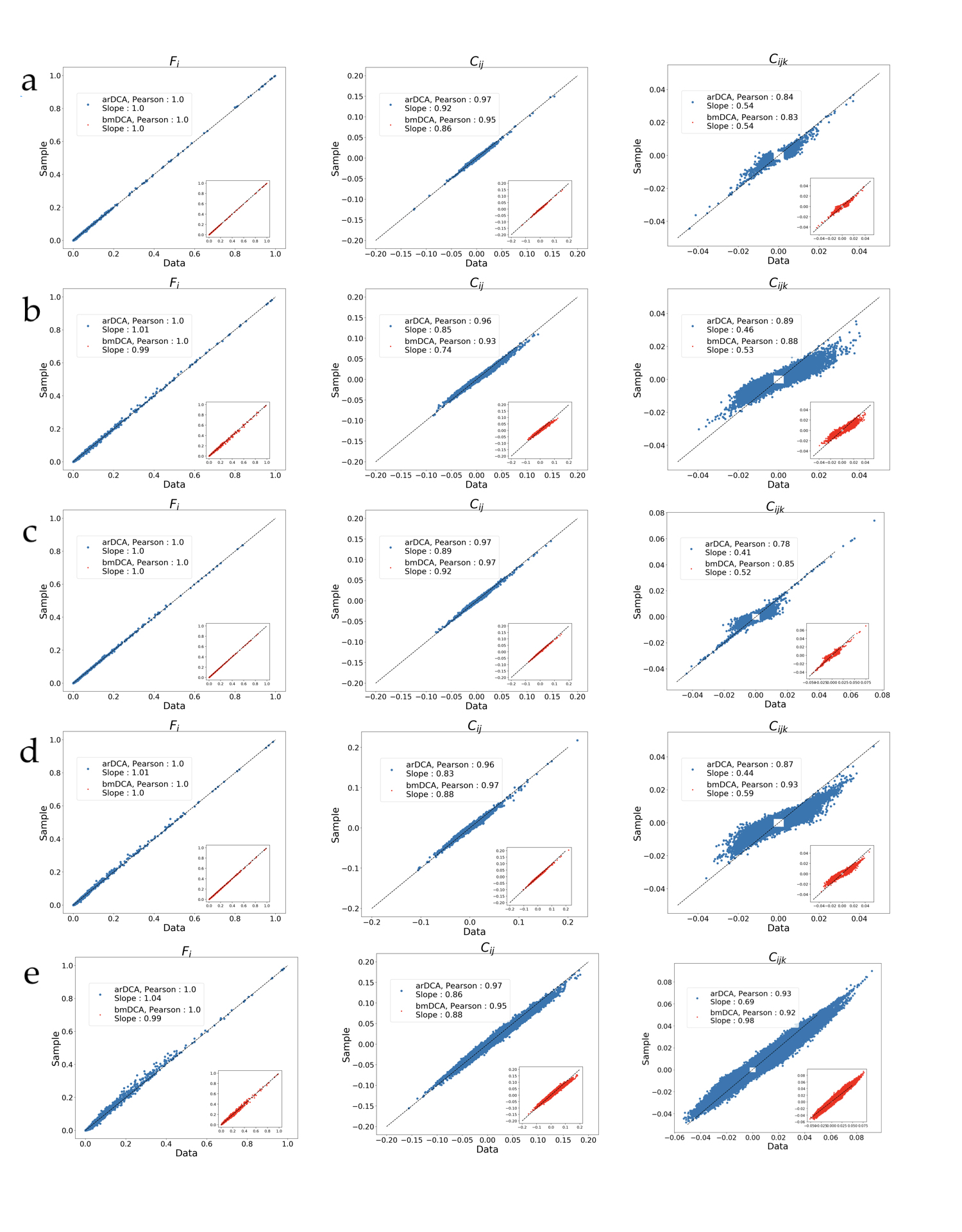}
\vskip-30pt
\caption{One-point frequencies and two- and three-point connected correlations, obtained from resampling the models (vertical axis) and from empirical data (horizontal axis) for the Pfam families PF00014 (a), PF00072 (b), PF00076 (c), PF0595 (d), and PF13354 (e).}
\label{freq}
\end{figure}


\subsection{Families used for mutational effects}

We show in Supplementary Table~\ref{tab:S1} the generative properties of the arDCA model for the $33$ families that are used for mutational effect predictions~\cite{riesselman2018deep,laine2019gemme}. The computational time of parameter learning on a standard laptop is also included.


    \begin{table}
    \begin{tabular}{|m{4cm}|m{2cm}|m{2cm}|m{2cm}|m{2cm}|m{3cm}|}
        \hline
         Family  & $L$  & $M$  & $M_{\rm eff}$ & Time (min) & Pearson's correlation of $C_{ij}$\\
         \hline
         YAP1 (WW domain) & 30 & 85299  & 5822 & 4 & 0.99\\
         \hline
         GAL1 (DNA-binding domain) & 62 & 20688 & 6435 & 5 & 0.98\\
         \hline
         Translation initiation factor IF1 & 69 & 9090 & 1310 & 2 & 0.98\\
         \hline
         RAL1 & 71 & 33026 & 6435 & 9 & 0.97\\
         \hline 
         BRCA1 (RING domain) & 75 & 39396 & 6585 & 12 & 0.96 \\
         \hline 
         UBE4B (U-box domain) & 75 & 16478 & 2941 & 5 & 0.98 \\
         \hline 
         Small ubiquitin-related modifier 1 & 76 & 21695 & 2669 & 5 & 0.99 \\
         \hline
         PABP & 79 & 246405 & 29045 & 77 & 0.97 \\
         \hline 
         PSD95 (PDZ factor) & 82 & 208112 & 7215 & 60 & 0.97 \\
         \hline
         Hepatitis C NS5A & 113 & 11423 & 55 & 4 & 1 \\
         \hline 
         SUMO conjugating enzyme UBC9 & 138 & 32486 & 4957 & 37 & 0.97 \\
         \hline
         Calmodulin-I & 139 & 36224 & 7196 & 30 & 0.94 \\
         \hline
         GTPase HRas & 164 & 84762 & 12506 & 130 & 0.98 \\
         \hline 
         Thiopurine S-Methyltransferase & 177 & 6688 & 2351 & 13 & 0.98 \\
         \hline 
         BRCA1 (BRCT domain) & 186 & 8391 & 2037 & 14 & 0.94 \\
         \hline 
         Thiamin Pyrophosphokinase 1 & 201 & 9966 & 3851 & 26 & 0.98 \\
         \hline 
         HSP90 (ATPase domain) & 218 & 23447 & 2847 & 40 & 0.98 \\
         \hline 
         Kanamycin kinase APH(3') -II & 226 & 29808 & 9658 & 60 & 0.96 \\
         \hline 
         TIM Barrel (T.thermophilus) & 236 & 23742 & 4869 & 98 & 0.97 \\
         \hline 
         TIM Barrel (T.Solfataricus) & 237 & 23743 & 4913 & 101 & 0.97 \\
         \hline 
         TIM Barrel (T.Maritima) & 239 & 23745 & 5001 & 103 & 0.97 \\
         \hline
         Aliphatic amide hydrolase & 247  & 76372  & 20145 & 340 & 0.97 \\
         \hline 
         Beta-lactamase & 252 & 14783 & 3818 & 60 & 0.97 \\
        \hline  
         Mitogen-activated protein kinase 1 & 288 & 65626 & 7322 & 400 &  0.98 \\
         \hline 
         PTEN & 304 & 8566 & 1119 & 43 & 0.96 \\
         \hline 
         DNA methylase HaeIII & 318 & 26513 & 11098 & 230 & 0.96 \\
         \hline
         Levoglucosan kinase & 364 & 12925 & 3638 & 160 & 0.98 \\
         \hline 
         Beta-glucosidase & 441 & 49471 & 8477  & 400 & 0.98 \\
         \hline
         Influenza Hemagglutin & 544 & 51000 & 62 & 500 & 0.99 \\
         \hline 
         HIV env (BF520) & 657 & 73441 & 305 &  800 & 0.98 \\
         \hline 
         Influenza polymerase PA subunit & 716 & 19611 & 9 & 518 & 1 \\
         \hline
        
    \end{tabular}
    \caption{MSA parameters, running time and Pearson correlation for $C_{ij}$ for
    the 32 families used for comparison with DMS data.}
    \label{tab:S1}
    \end{table}

\subsection{Comparison of mutational predictions of DeepSequence and arDCA}

\begin{figure}
    \centering
    \includegraphics[scale = 0.2]{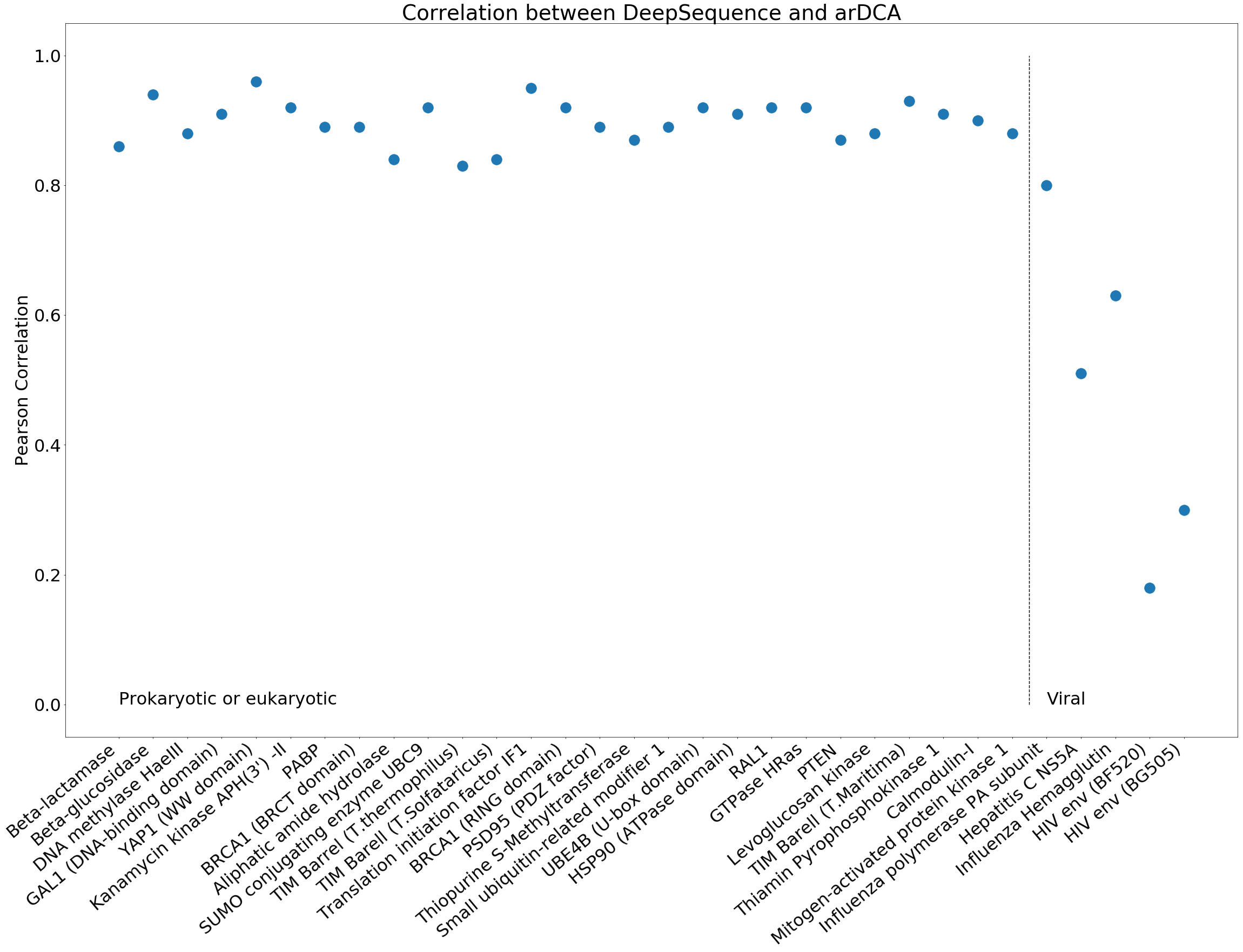}
    \caption{Pearson correlations for mutational predictions of arDCA and DeepSequence across all protein families studied.}
    \label{mut_effects_DeepS_vs_arDCA}
\end{figure}

\clearpage

\section{Contact prediction}
\label{app:cont}

Once the effective couplings are calculated, the standard procedure of DCA is applied~\cite{cocco2018inverse}. First,  each pair of amino acids is assigned an interaction score given by the Frobenius norm: 
\begin{equation}
   F_{ij} =  \| \vect{J}_{ij}\| ^2 = \sqrt{\sum_{a,b = 1}^{20}J_{ij}(a,b)^2} \ .
\end{equation}
Note that the gap state ($q = 21$) is not taken into account in the norm.
Because of overparametrization caused by the non-independence of the empirical frequencies,
both the Potts model and autoregressive models are invariant under some gauge transformations, that is different sets of parameters give the same probability. On the other side, the Frobenius norm is not invariant by gauge transformation, so a gauge choice is needed. The zero-sum gauge was found to be the gauge that minimizes the Frobenius norm; in other words, this gauge choice includes in the couplings only the information that cannot be treated by fields. Note that the zero-sum gauge is also the gauge of the standard Ising model. The equations characterizing the zero-sum gauge are: $\sum_{a} h_i(a) = \sum_{a} J_{ij}(a,b) = \sum_{b} J_{ij}(a,b) =  0$. Finally, the so-called average product correction (APC) is applied on the Frobenius scores, because it was empirically shown to improve contact prediction: $F_{ij}
^{APC} = F_{ij}- \frac{F_{.j}F_{i.}}{F_{..}}$ where the dot represents the average with respect to the index.

\subsubsection{Results across families}
\paragraph{PPV --}To test the contact prediction, a distance $d(i,j)$ between each heavy atoms in the amino acids was extracted from the crystal structures present in the PDB  database. Sites with atoms at a distance $<8$ Å were considered in contact. Note that $8$ Å is too large to be a true contact, but since we are looking for consensus contacts in the family and there is variability from protein to protein, this definition has become standard. Coherently with the literature standard, a minimal separation of $|i-j| \geq 5$ along the protein chain was imposed in order to consider only non-trivial contacts corresponding to sites that are not close in the chain.
Pairs $ij$ are ranked according to the APC-corrected Frobenius norm, as defined in ~\ref{app:cont}.
Supplementary Figure~\ref{PPV} shows the positive predictive value as a function of the number of predicted non-trivial contacts for the arDCA model (blue) and bmDCA (orange). The Positive Predicted Value (PPV)  is the fraction of true contacts among the first $n$ predictions, corresponding to the $n$ highest scores.

\begin{figure}
    \centering
    \includegraphics[scale = 0.42]{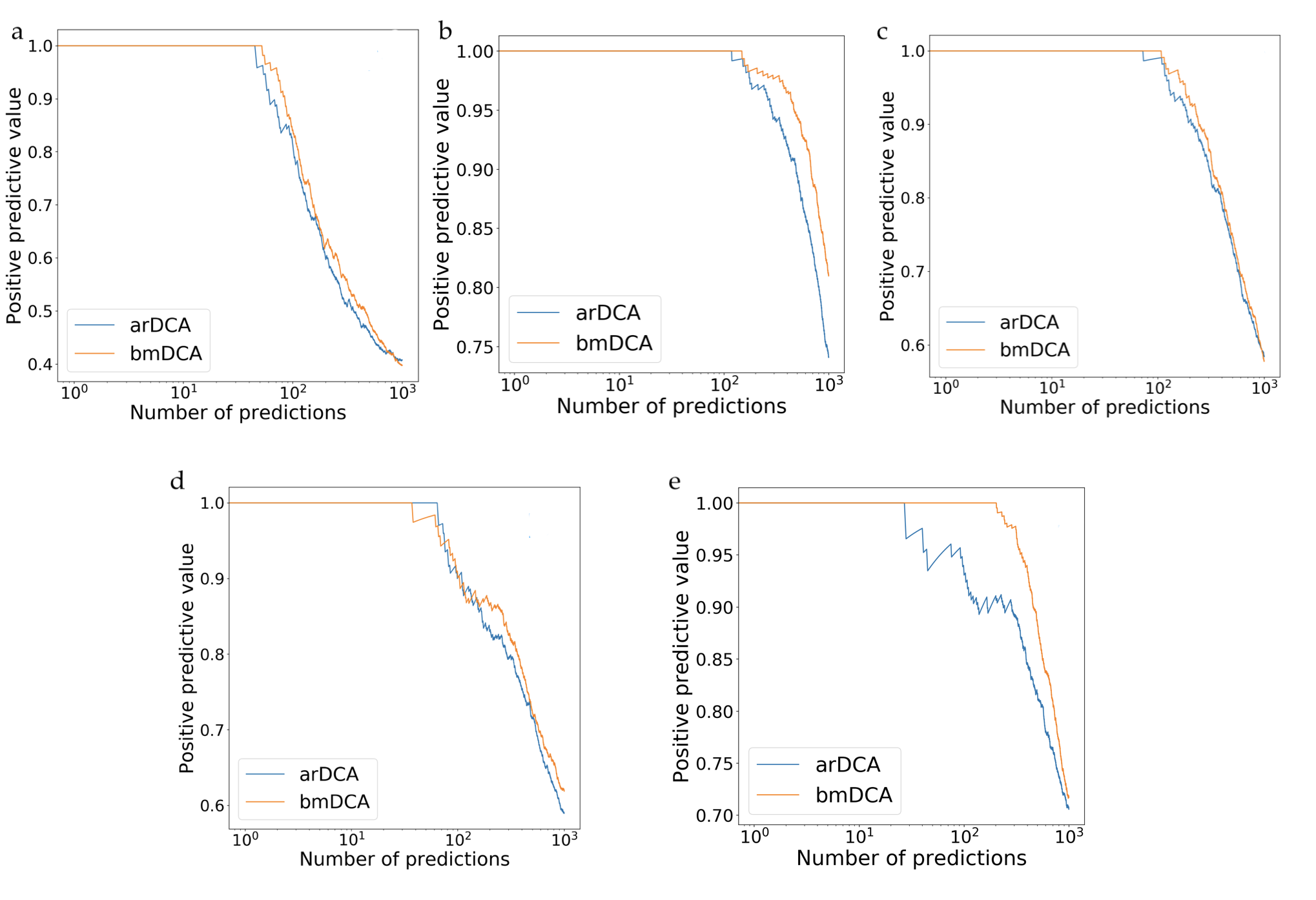}
    \caption{PPV curves for the Pfam families PF00014 (a), PF00072 (b), PF00076 (c), PF0595 (d), and PF13354 (e) obtained with arDCA and bmDCA.}
    \label{PPV}
\end{figure}

\begin{figure}
    \centering
    \includegraphics[scale = 0.5]{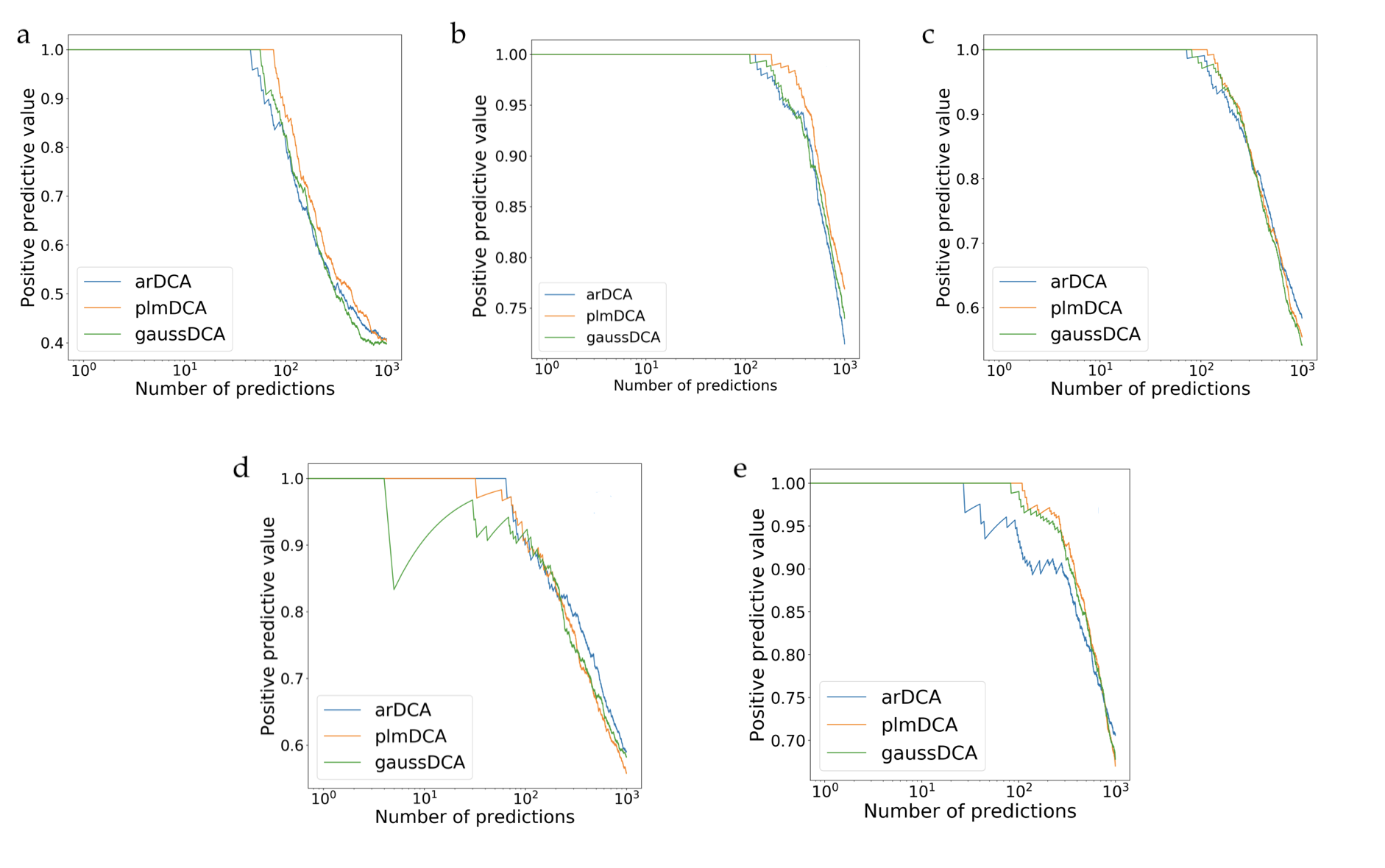}
    \caption{PPV curves for the Pfam families PF00014 (a), PF00072 (b), PF00076 (c), PF0595 (d), and PF13354 (e) obtained with arDCA, plmDCA and GaussDCA (aka mfDCA).}
    \label{PPV2}
\end{figure}

\paragraph{Contact map --}Supplementary Figure~\ref{contact map} show the contact maps of the arDCA and bmDCA models. The black crosses represent the true contact map with a threshold of $8$ Å. The blue dots are the true positive predictions and the red ones are the false positive considering the $2L$ top predictions of non-trivial contacts.

\begin{figure}
    \centering
    \includegraphics[scale = 0.55]{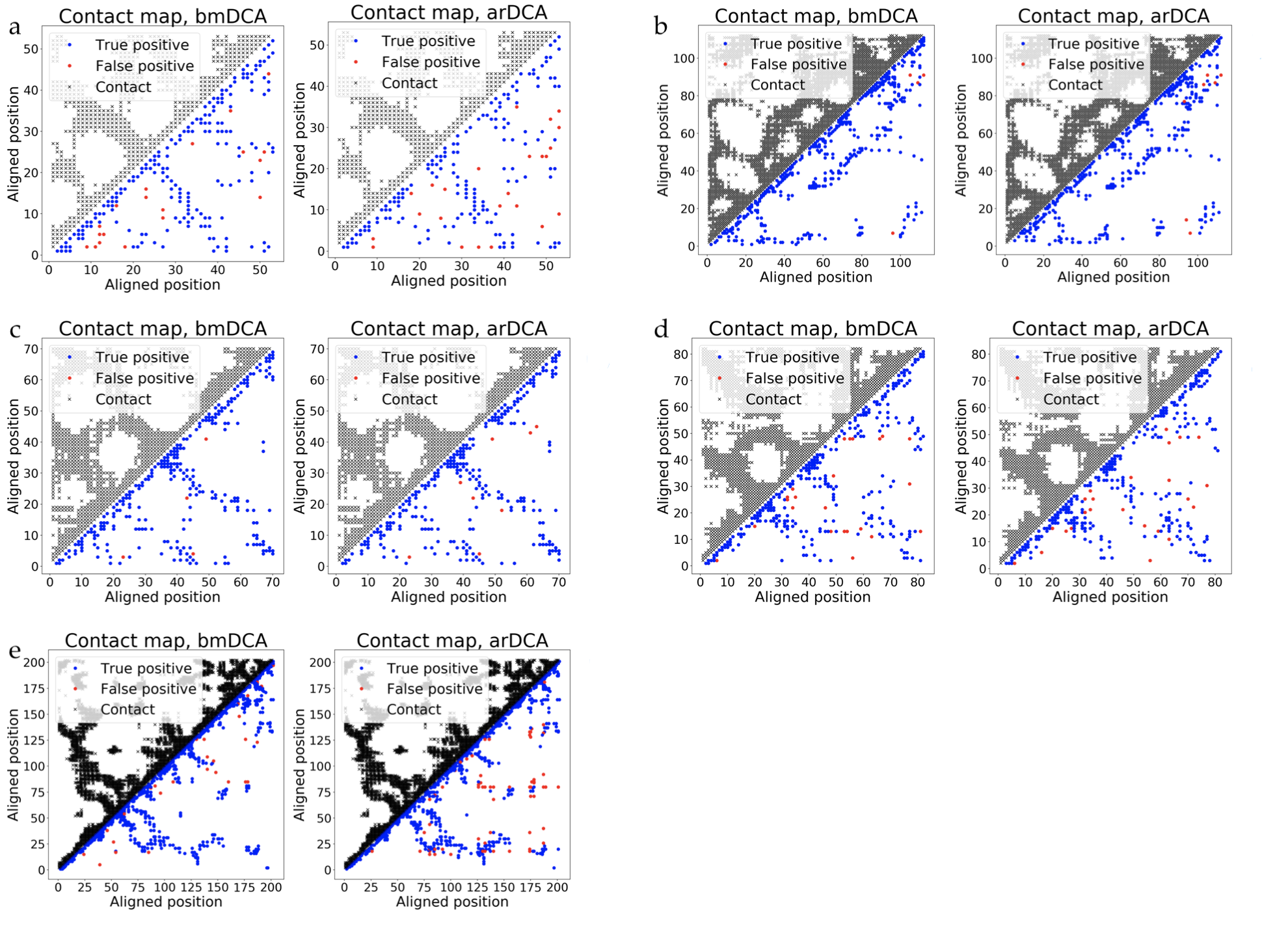}
    \caption{Contact maps for the Pfam families PF00014 (a), PF00072 (b), PF00076 (c), PF0595 (d), and PF13354 (e).}
    \label{contact map}
\end{figure}

\clearpage

\begin{figure}
    \centering
    \includegraphics[scale = 0.55]{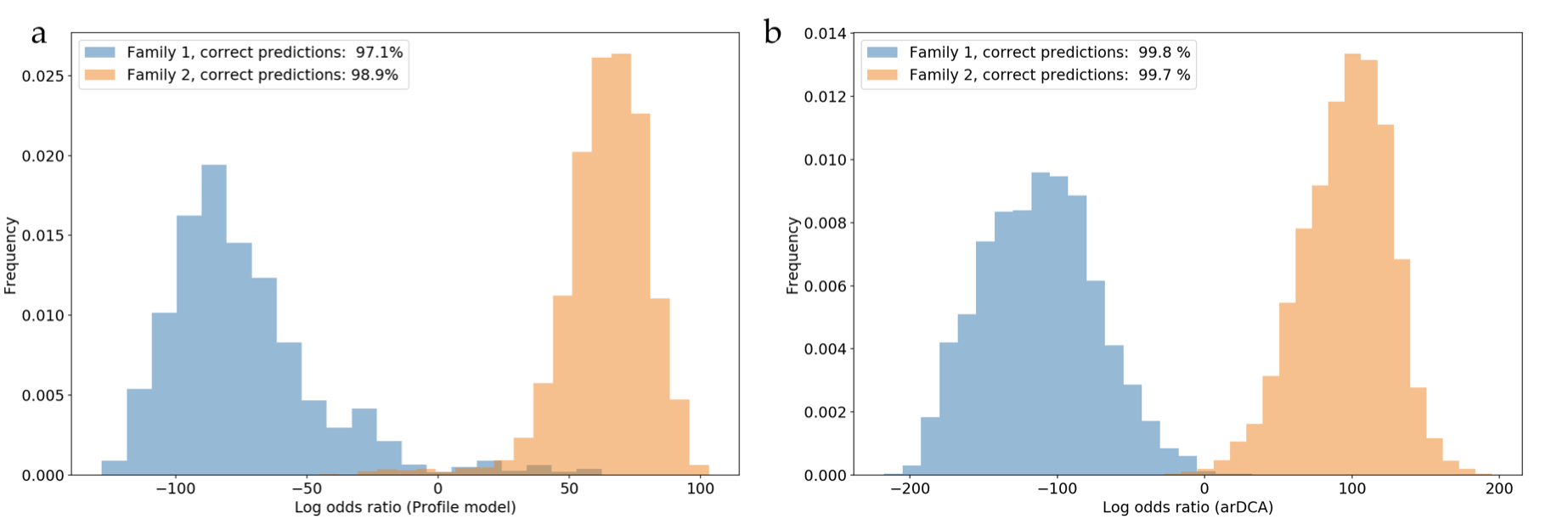}
    \caption{Histograms of log-odds rations $\log\{P_1(seq)/P_2(seq)\}$ for subfamily specific profile (panel a) and arDCA (panel b) models, for the two sub-families of the response-regulator domain family (PF00072) identified by the coexistence with one of two DNA-binding domains Trans\_reg\_C (PF00486) or GerE (PF00196). Models are learned on randomly extracted training sets of 6000 sequences per sub-family, histograms show the test sets (blue for coexistence with PF00486, orange with PF00196). The part of the blue (resp. orange) histograms with negative (resp. positive) values corresponds to correct sub-family annotations via the log-odds ratio; the fraction of correct predictions for each subfamily and each model is indicated in the legend.}
\end{figure}

\begin{figure}
    \centering
    \includegraphics[scale = 0.4]{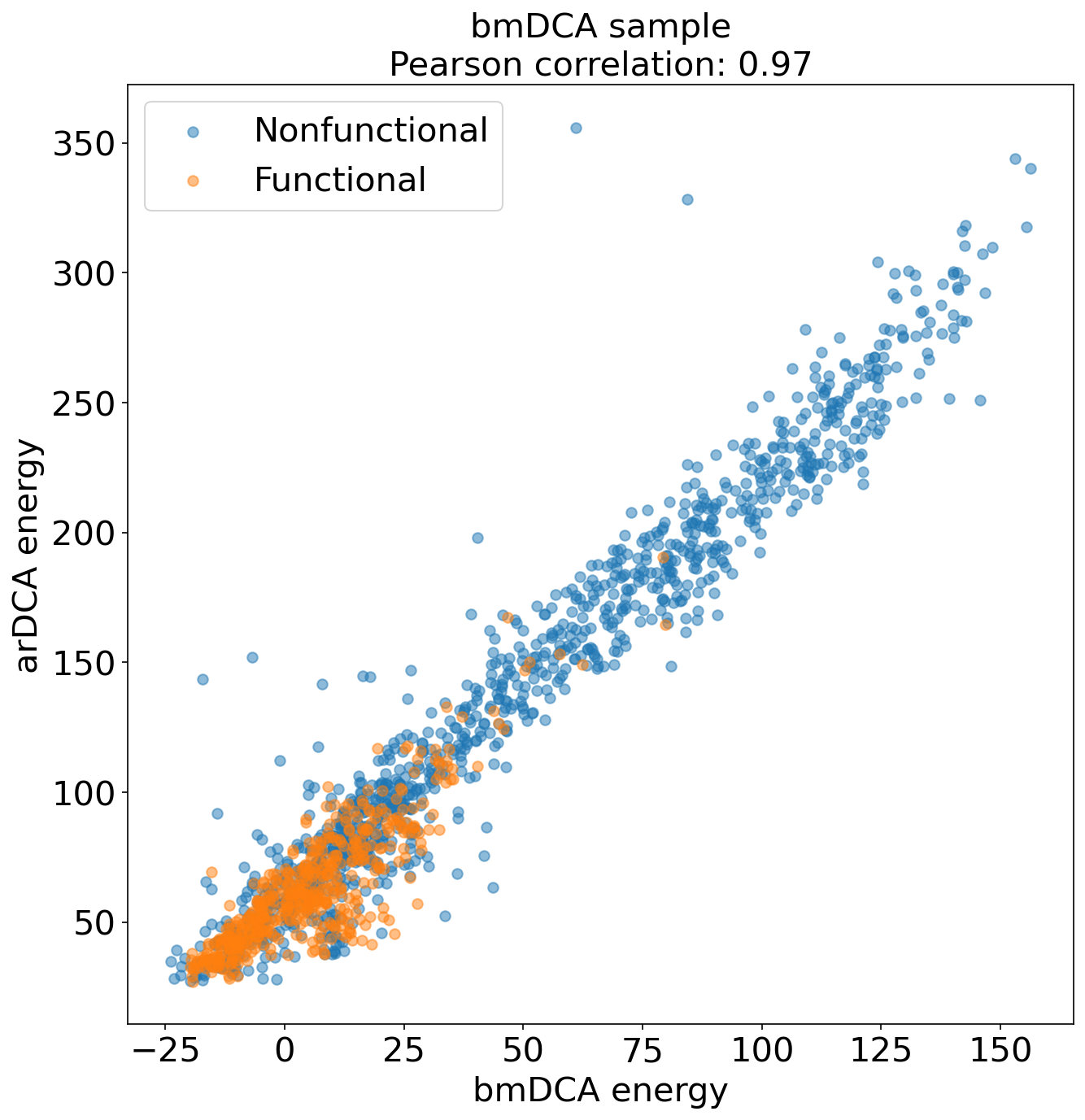}
    \caption{Scatter plot of statistical energies of bmDCA and arDCA for the artificial chorismate mutase enzymes enzymes designed in \cite{russ2020evolution}; the bmDCA energies are the ones published along with the references. The coloring allows to distinguish experimentally verified functional (red) and non-functional (blue) sequences. Both energies are highly correlated (Pearson correlation 0.97), and functional sequences are found at low energies only, in perfect agreement with \cite{russ2020evolution}.  }
    \label{contactmap2}
\end{figure}

\begin{figure}
    \centering
    \includegraphics[scale = 0.55]{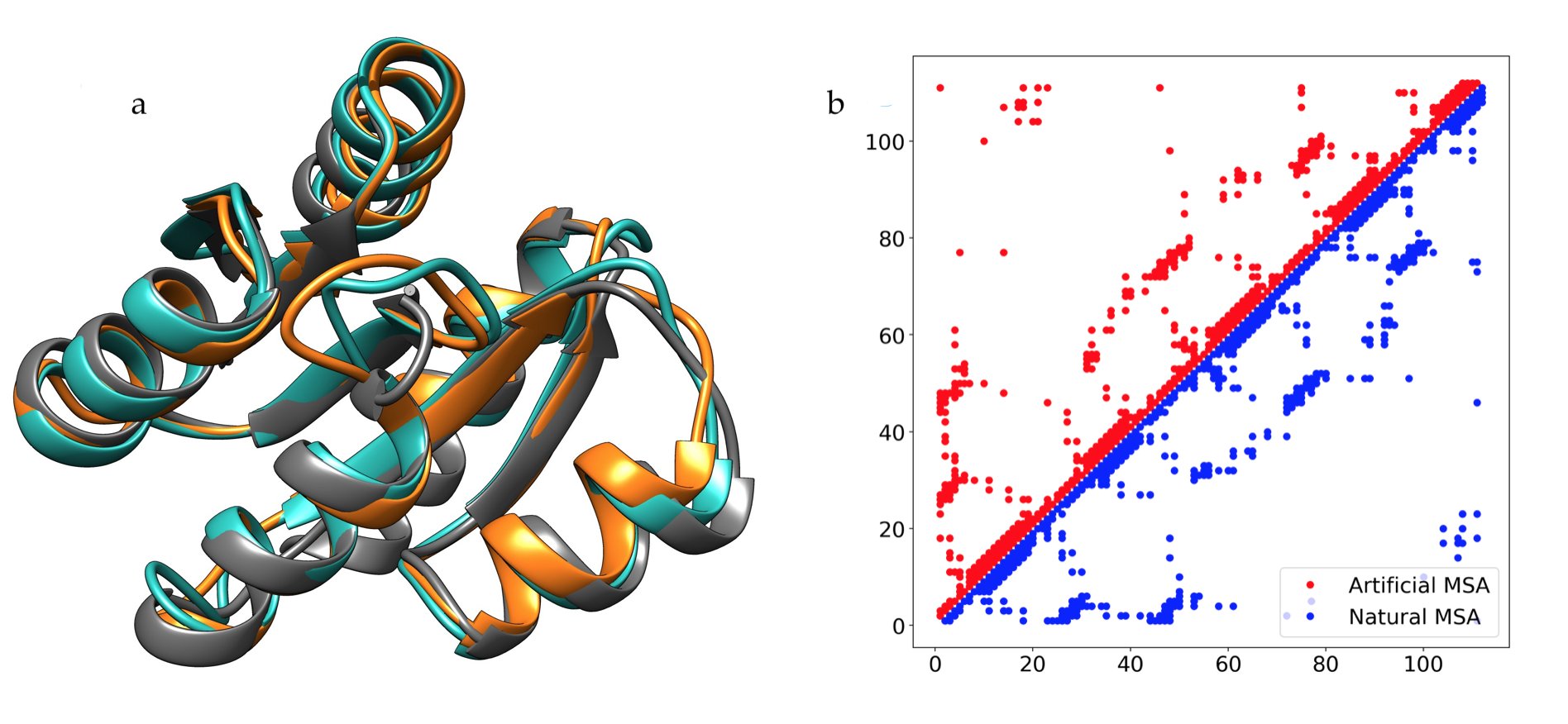}
    \caption{Panel a: Comparison of an exemplary PDB structure of the PF00072 family (PDB ID 1nxs \cite{bent2004crystal}, grey) with trRosetta predictions for small MSA of 10 arDCA-generated sequences (turquoise, RMSD 1.96\AA, and 0.91\AA~over 94/106 residues with 2\AA) and of 10 natural sequences (orange, RMSD 1.74\AA, and 0.91\AA~over 90/108 residues with 2\AA). Panel b shows the contact maps for the two trRosetta predictions.}
\end{figure}

\clearpage

\section{Two-layer autoregressive models}

\begin{figure}[t]
      \centering
      \includegraphics[scale = 0.6]{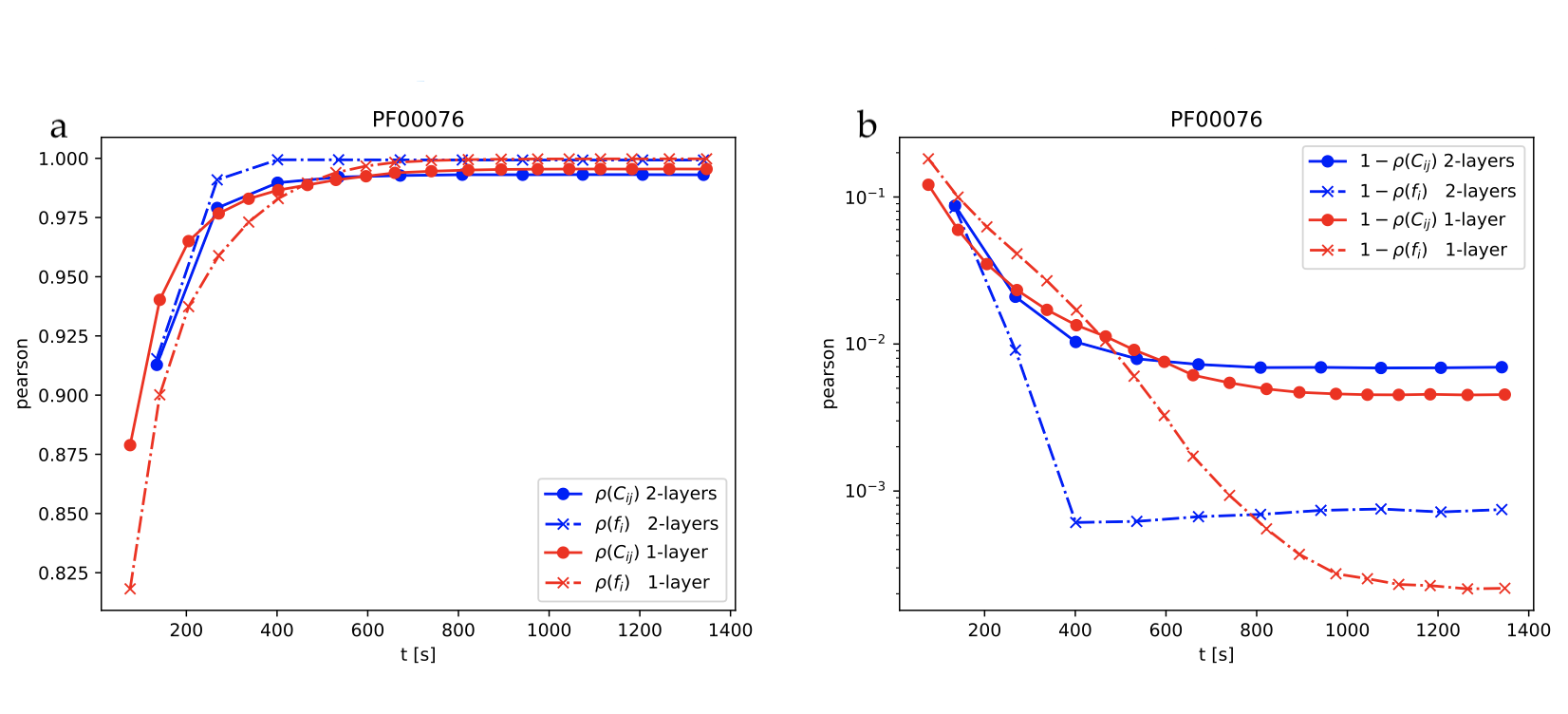}
      \caption{Evolution of the Pearson correlation (for one- and two-point statistics) during the learning for the family PF00076. The computational time is given in seconds. Both figures compare the one-layer (red) and the two-layer (blue) models. Figure a shows the Pearson correlation while figure b shows one minus the Pearson correlation (in semilog scale). }
      \label{2_layers}
\end{figure}

Due to the very simple structure of the one-layer arDCA model, one might ask whether a more complicated and flexible model could perform better. To address this question, we considered an arDCA model for the family PF00076, but with two layers instead of one. As an exploratory step, we just considered very simple two-layer architectures: each conditional probability $P(a_i | a_{i-1},\dots,a_{1})$, $i \in 2,\dots,L$  is modeled in terms of a dense input node of size $kq \times (i-1)q$ for (input amino acid sequences are one-hot-encoded) with non a linear activation function $\sigma$, while the second layer is again a dense node of size $q \times k q $ concatenated with a {\it softmax} to get a probability as final output:
\begin{equation}
\tilde{P}_{i}^{(2-layer)}(a_1,\dots,a_{i-1})=\mathrm{softmax}\left(W_{2}^{[q,kq]}\cdot\sigma\left(W_{1}^{[kq,(i-1)q]}\cdot\vec{a}+\vec{b}^{[kq]}_1 \right)+\vec{b}_{2}^{[q]}\right)\quad,\quad i\in \{2,\dots,L\}\quad,
\end{equation}    
where $W_{1,2}^{[l,m]}$ are parameter matrices of size $l\times m$ and $\vec{b}_{1,2}^{[l]}$ are vectors of parameters (biases) of size $l$ that are optimized in the training step. We tried different values of $k$, and different types of activation functions $\sigma$ and we opted for $k = 5$ and $\sigma = \mathrm{ReLU}$. 

Supplementary Figure~\ref{2_layers} shows that increasing the complexity of the model does not improve the ability to reproduce the statistics of the data. In the specific case of the family PF00076, the one-layer model is even marginally better, both for the one-point and two-point statistics. Moreover, the computational time is comparable in the two cases, therefore using a two-layer model gives neither an advantage on the generative qualities, nor on the computational time.

\end{widetext}


\bibliography{refs_ar}
\bibliographystyle{apsrev4-1}

\end{document}